\documentclass[aps,prd,reprint,nofootinbib,longbibliography,floatfix]{revtex4-1}

\usepackage[english]{babel}
\usepackage{hyphenat}
\usepackage{graphicx}
\usepackage{array}
\usepackage{dcolumn}
\usepackage{bm}
\usepackage{amssymb}
\usepackage{amsmath}
\usepackage{amsfonts}
\usepackage{amsbsy}
\usepackage{latexsym}

\usepackage[makeroom]{cancel}
\usepackage[breaklinks,colorlinks,citecolor=blue,urlcolor=blue,linkcolor=blue]{hyperref}

\begin{document}

\title{ Lorentz invariant relative velocity and relativistic binary  collisions}

\author{Mirco Cannoni}
\email{mirco.cannoni@dfa.uhu.es}

\affiliation{Departamento de F\'isica Aplicada, Facultad de Ciencias Experimentales, Universidad de Huelva, 21071 Huelva, Spain}

\date{May 2, 2016 }

\begin{abstract}

This article reviews  the concept of Lorentz invariant relative velocity 
that is often misunderstood or unknown in high energy physics literature. 
The properties of the relative velocity allow to formulate 
the invariant flux and cross section
without recurring to non--physical velocities or any assumption about the reference frame. 
Applications  such as  the  luminosity of a collider, 
the use as kinematic variable, and 
the statistical theory of collisions in a relativistic classical gas are reviewed.
It is emphasized how the hyperbolic properties of the velocity space explain 
the peculiarities of relativistic scattering.

\end{abstract}

\maketitle
\tableofcontents

\section{Introduction}
\label{sec:intro}

In every book on relativistic quantum field theory and particle physics
a section or an appendix is necessarily devoted to the formulation of the 
Lorentz invariant cross section. To fix the ideas we consider the total cross section 
for binary collisions $1+2 \to f$ where $f$ is a final state with 2 or more particles.

The basic quantity that is  measured in experiments
is the reaction rate, that is the number of events with the final state $f$ 
per unit volume per unit of time, 
\begin{flalign}
\mathcal{R}_f=\frac{dN_f}{dVdt}=\frac{dN_f}{d^4 x}.
\end{flalign}
The value of the reaction rate is proportional to the
number density of particles $n_1$ and  $n_2$ that approach each other with a certain relative 
velocity, for example an incident beam on a fixed target 
or two colliding beams. This is the so--called initial (or incident) flux $F$. 
The physical quantity that gives the intrinsic quantum probability for a transition 
independent from the details of the initial state
is the cross section defined as the ratio $\sigma={\mathcal{R}_f}/{F}$.

In nonrelativistic scattering the initial flux is given by 
\begin{equation}
F_{\text{nr}}=n_1 n_2 v_r,
\label{flux_nonrel}
\end{equation}
where 
\begin{equation}
v_r=|\boldsymbol{v}_1 - \boldsymbol{v}_2|,
\label{vrel_nonrel}
\end{equation}
is the nonrelativistic relative velocity.\footnote{When explicitly written, we 
assume that the velocities are given in the laboratory frame. 
Although sometimes used as synonymous, we here 
distinguish the laboratory from the rest frame of massive particles.}

Since the number of events $N_f$ does not depend on the reference frame and $d^4 x$ is 
invariant, it is essential that  the initial flux $F$ and the cross section $\sigma$  
are invariant under proper Lorentz transformations such that their product $\mathcal{R}_f$ is 
invariant.
The expressions 
(\ref{flux_nonrel}) and (\ref{vrel_nonrel}) are not invariant under Lorentz transformations
and are not valid in the relativistic framework. 

We start this review, Section \ref{sec:1}, with a critical discussion of how the 
relativistic flux is presented in textbooks.
This will lead us to review in Section \ref{sec:v_rel} 
the properties  of the invariant relative velocity and 
to discuss a simple Lorentz invariant definition of the flux in Section \ref{sec:flux}. 
We then review some  important applications to the theory of relativistic scattering: 
the luminosity of a collider in Section 
\ref{sec:luminosity}, 
the use of the invariant relative velocity as 
kinematic variable in Section \ref{sec:kin_var},
and the theory of collisions in a  relativistic gas in Section 
\ref{sec:gas}. 
Overall, we shall emphasize how the hyperbolic nature of the relativistic velocity space
manifests itself in concrete physical applications.

\section{The definition of invariant cross section: three problems and a solution}
\label{sec:1}

In  quantum field theory the invariant rate corresponds to the probability per unit time per 
unit volume of a single transition $i \to  f$, 
$\mathcal{R}^\text{th}_\text{fi}=\frac{dP_\text{fi}}{dVdt}$ 
and is given by
\begin{flalign}
\mathcal{R}^\text{th}_\text{fi}=\int 
\overline{|\mathcal{M}_\text{fi}|^2}
(2\pi)^4\delta^4(P_i - P_f)
\prod_{j=3}^{n} \frac{d^3 \boldsymbol{p}_j}{(2\pi)^3 2E_j}.
\label{rate_def_theo}
\end{flalign}
Equation (\ref{rate_def_theo}) refers to the scattering of unpolarized particles, 
hence the square of the amplitude is
summed over the final spins and averaged over the initial spins.  
With  $P_i$ and $P_f$ we indicate  the total 4-momentum
of the initial and final states, respectively.

In (\ref{rate_def_theo}) we use the normalization of one--particle states $\langle p|p' 
\rangle=(2\pi)^3 2E \delta^3 (\boldsymbol{p}-\boldsymbol{p}')$ 
both for bosons and fermions, which corresponds to  `$2E$ particles per unit volume' instead of `one 
particle per unit volume'.
The densities appearing in the flux according to this convention are thus $n=2E$.
We shall see that for an experimentalist or in the case of a gas
of particles, the densities are something more concrete.

Although in the Introduction we argued that Eq.~(\ref{flux_nonrel}) 
is not valid in the relativistic 
framework, the starting point of quantum field theory and particle physics textbooks 
is nonetheless the nonrelativistic expression 
\begin{flalign}
F=n_1 n_2 \rvert \boldsymbol{v}_1 - \boldsymbol{v}_2 \rvert, 
\label{flux_books}
\end{flalign}
where  velocities $\boldsymbol{v}_1$ and  $\boldsymbol{v}_2$ are required to be 
\textit{collinear}. This includes as particular cases the expression of the flux in the 
rest frame of massive particles and in the center of momentum frame. We shall discuss in a moment
this assumption. Let us proceed and 
use $\boldsymbol{v}=\boldsymbol{p}/E$ to rewrite Eq.~(\ref{flux_books}) as 
\begin{flalign}
n_1 n_2\Big\rvert \frac{E_2 \boldsymbol{p}_1- E_1 \boldsymbol{p}_2}{E_1 E_2} \Big\rvert.
\nonumber
\end{flalign}
Employing $E^2=m^2 +\boldsymbol{p}^2$ we find
\begin{flalign}
&|E_2 \boldsymbol{p}_1 - E_1 \boldsymbol{p}_2|\cr
&=[
m^2_2 \boldsymbol{p}_1^2 +m^2_2 \boldsymbol{p}_2^2
-2E_1 E_2 \boldsymbol{p}_1\cdot \boldsymbol{p}_2 
+2 \boldsymbol{p}_1^2 \boldsymbol{p}_2^2
]^{1/2}.
\label{exp1}
\end{flalign}
Now  we note  that\footnote{
Four-vectors are indicated as $a=(a^0 ,\boldsymbol{a})$ and the Minkowski scalar product
as $a\cdot b=a^0b^0-\boldsymbol{a}\cdot \boldsymbol{b}$. Natural units $\hbar=c=k_B=1$ are used.} 
\begin{flalign}
&\mathcal{F}\equiv \sqrt{(p_1 \cdot p_2)^2 -m^2_1 m^2_2}=\nonumber\\
&[
m^2_2 \boldsymbol{p}_1^2 +m^2_2 \boldsymbol{p}_2^2
-2E_1 E_2 \boldsymbol{p}_1\cdot \boldsymbol{p}_2
+ \boldsymbol{p}_1^2 \boldsymbol{p}_2^2 
+(\boldsymbol{p}_1 \cdot \boldsymbol{p}_2)^2
]^{1/2},
\label{exp2}
\end{flalign}
where we used again the relativistic energy formula.
If the velocities are collinear, the last two terms in (\ref{exp2})
are equal and sum to give the last term in (\ref{exp1}). In this case  the expressions
(\ref{exp1}) and (\ref{exp2}) coincide and 
the flux (\ref{flux_books}) becomes
\begin{flalign}
F=\frac{n_1 n_2}{E_1 E_2} \mathcal{F}.
\label{flux_standard_n1n2}
\end{flalign}
The product of the number densities is $n_1 n_2=4 E_1 E_2$, 
hence 
\begin{flalign}
F=4\mathcal{F}.
\label{flux_standard}
\end{flalign}
This is the most popular form of the flux in particle physics also reported in the review of the 
Particle Data Group \cite{Agashe:2014kda}. 

\subsection{Why collinear velocities?}

If we start from Eq.~(\ref{flux_books}), the assumption of collinearity is necessary because, 
taking the velocities along the $z$ axis for 
example, the expression 
$ E_1 E_2 |v_{1_z}-v_{2_z}|=|E_2 p_{1_z}-E_1 p_{2_z}|$ 
is at least invariant under boosts along the $z$ axis, even if not under a general 
Lorentz transformation. This corresponds to the simplification of the last terms in Eq.~(\ref{exp2}).

The final expression (\ref{flux_standard}) is anyway a Lorentz scalar, hence also valid 
for noncollinear velocities\footnote{
In many textbooks, for example \cite{bjorken1964relativistic}, \cite{Halzen:1984mc}, 
\cite{Brown:1992db},~\cite{Kaku:1993ym}, 
\cite{greiner2008quantum},~\cite{Tully:2011zz}
it is incorrectly stated that Eq.~(\ref{flux_standard}) is valid only for collinear velocities.
The motivation is that the cross section
must be invariant under  boosts along the perpendicular direction in order to transform 
as an area transverse to the beams direction, see also \cite{Peskin:1995ev}, \cite{Zee:2003mt}.
This argument is irrelevant for the definition of the theoretical quantum cross section, 
and we shall see in Section \ref{sec:luminosity}, when discussing luminosity of a collider, 
that collisions with real beams are
almost never collinear, thus full Lorentz invariance is necessary. }
as can be easily seen in the following  way.
Dividing Eq.~(\ref{exp2}) by $E_1 E_2$, the right hand side can be written as
\begin{flalign}
& 
\left[\left(\frac{\boldsymbol{p}_1}{E_1}-\frac{\boldsymbol{p}_2}{E_2}\right)^2
+\left(\frac{\boldsymbol{p}_1}{E_1}\cdot 
\frac{\boldsymbol{p}_2}{E_2}\right)^2-\left(\frac{\boldsymbol{p}_1}{E_1}\right)^2 
\left(\frac{\boldsymbol{p}_2}{E_2}\right)^2\right]^{1/2}
\nonumber\\
&\;=
[(\boldsymbol{v}_1 - \boldsymbol{v}_2)^2+(\boldsymbol{v}_1 \cdot \boldsymbol{v}_2)^2 
-\boldsymbol{v}^2_1 \boldsymbol{v}^2_2]^{1/2}.
\label{9bis}
\end{flalign}
Using  the vector identity 
\begin{flalign}
(\boldsymbol{a} \times \boldsymbol{b})^2=\boldsymbol{a}^2 \boldsymbol{b}^2 -( \boldsymbol{a}\cdot \boldsymbol{b})^2
\label{vector_identity}
\end{flalign}
in (\ref{9bis}), it follows 
\begin{flalign}
\frac{\mathcal{F}}{E_1 E_2}=
\sqrt{(\boldsymbol{v}_1 - \boldsymbol{v}_2)^2 - 
{(\boldsymbol{v}_1 \times \boldsymbol{v}_2)^2}}\equiv\bar{v}.
\label{v_moller_v} 
\end{flalign}
Therefore for every $\boldsymbol{v}_1$ and $\boldsymbol{v}_2$ we can write
\begin{flalign}
F&=n_1 n_2 \sqrt{(\boldsymbol{v}_1 - \boldsymbol{v}_2)^2 - 
(\boldsymbol{v}_1 \times \boldsymbol{v}_2)^2}.
\label{flux_standard_n1n2_Moller} 
\end{flalign}
The collinear flux (\ref{flux_books}) is just a particular case of 
the invariant expression (\ref{flux_standard_n1n2_Moller}) when $\boldsymbol{v}_1 \times 
\boldsymbol{v}_2=0$.

\subsubsection*{The M\o{}ller's flux}

Formula~(\ref{flux_standard_n1n2_Moller}) was proposed
a long time ago in~\cite{Moller1945}.
In this paper M\o{}ller  notes that textbooks define (at that time) the cross section in the 
center of momentum frame with the flux given by Eq. (\ref{flux_books}) (hence not differently from 
today). His purpose is to find a general expression 
valid also for non collinear velocities.\footnote{
The invariant M\o{}ller flux  
was reported in some influential books such as \cite{jauch1955theory}, 
\cite{goldberger1964collision} (and \cite{Landau:1982dva} as we shall discuss at length)
but then disappeared from quantum field theory and particle physics 
literature, replaced by the 'dogma' of collinearity.
}

M\o{}ller does not derive, but affirms 
that such an expression is given by Eq.~(\ref{flux_standard_n1n2_Moller}) 
and then prove the invariance. 
In order to do that, he writes the squared root in (\ref{flux_standard_n1n2_Moller})
as $\frac{B}{E_1 E_2}$, with
\begin{flalign}
B=\sqrt{(E_2 \boldsymbol{p}_1- E_1 \boldsymbol{p}_2)^2-
(\boldsymbol{p}_1\times \boldsymbol{p}_2)^2}.
\nonumber
\end{flalign}
The squared expressions under the squared root are antisymmetric in the indices 1 and 2.
M\o{}ller thus introduces the antisymmetric tensor
\[
A^{\mu \nu}=p^\mu_1 p^\nu_2 -p^\nu_1 p^\mu_2,
\]
and it is easy to see that
\begin{flalign}
B=\sqrt{-\frac{1}{2} A^{\mu \nu} A_{\mu \nu}} 
\nonumber
\end{flalign}
is a scalar and coincides with $\mathcal{F}$.  
The flux (\ref{flux_standard_n1n2_Moller}) is then rewritten as
\begin{flalign}
F=\frac{n_1 n_2}{E_1 E_2} B.
\label{moller_flux_with_B}
\end{flalign}
Since energy is the time component of the 4-momentum
and  number density is the time component of the 4-current
\begin{flalign}
J=(n, n \boldsymbol{v})=n^0(\gamma,\gamma \boldsymbol{v})=n^0 u,
\label{4-current}
\end{flalign}
where $n^0$ is the number density in rest frame and $u$
the 4-velocity $u=\gamma(1,\boldsymbol{v})$, $u^2=1$,
$n_i$ and $E_i$ transform in the same way under Lorentz transformations.
The ratio $\frac{n_1 n_2}{E_1 E_2}$ is then a Lorentz invariant quantity. 
The flux is  invariant because is the   product of two scalar quantities.

Anyway, neither the derivation we have given, nor the M\o{}ller's paper clarifies 
how to derive expression (\ref{flux_standard_n1n2_Moller}) from first principles 
without starting from (\ref{flux_books}) with collinear velocities.

\subsection{`Who' is the relative velocity?}

It is important to remark that 
M\o{}ller  puts the product $E_1 E_2$ in ratio with the product of densities $n_1 n_2$, 
not with the scalar $B$.
He does not introduce  any `M\o{}ller velocity' or
call relative velocity the quantity $\bar{v}$ defined in (\ref{v_moller_v}).

The misleading identification of $\bar{v}=\mathcal{F}/E_1 E_2$ with a relative velocity or a 
`M\o{}ller velocity' is posterior,
probably suggested by the deceptive similarity of $F=n_1 n_2 \bar{v}$ with the nonrelativistic 
expression $F_\text{nr}=n_1 n_2 v_r$. 

As a matter of fact, in particle physics literature, some authors consider 
the quantity $\bar{v}$ as the relative velocity 
that generalizes the nonrelativistic expression $v_r$, for others the 
relative velocity is $v_r$.
In statistical physics literature~\cite{groot1980relativistic} and in dark matter 
literature~\cite{Gondolo:1990dk},  
$\bar{v}$ is called M\o{}ller velocity to distinguish it from the relative velocity,
which is considered to be given by $v_r$ in any case.

The above  identification is unfortunate because albeit the product $E_1 E_2 \bar{v}$ is a scalar, 
$\bar{v}$ by itself is not invariant.
Even worse, for many 
configurations and magnitudes of the two velocities, $v_r$ and $\bar{v}$ take values larger than
the velocity of light, both for massive and massless particles.
For example, in the center of momentum frame of two particles with equal masses, we have $v_r 
=\bar{v}=2 v_*$ and for $v_* > 0.5$ the `relative velocity' is superluminal.

Explicitly or tacitly, in high energy physics literature
is an accepted fact that the relative 
velocity of two particles can be larger than the velocity of light.\footnote{
Weinberg 
considers $\bar{v}$ as the relative velocity, called $u_\alpha$ in \cite{Weinberg:1995mt}, 
but, in evaluating the flux in the center of momentum frame, notes: "However, in this frame 
$u_\alpha$ is not
really a physical velocity; (...) for extremely relativistic 
particles, it can take values as large as 2".
}

In reality this is a macroscopic violation of the principles of relativity.
Fock~\cite{fock1969theory} expresses the point with the clearest words:
\begin{quotation}
In pre-relativistic mechanics the relative velocity of two bodies was defined as the 
difference of their velocities. Let the velocities of two bodies, both measured in the same frame of 
reference, be $\bf{u}$ and  $\bf{v}$ respectively. Then the velocity of the second body relative to 
the first used to be defined as $\bf{w}=\bf{v}-\bf{u}$. This definition is invariant with respect to 
Galileo transformations but not Lorentz transformations. Therefore it is not suitable in the Theory 
of Relativity and must be replaced by another. The fact that $\bf{w}=\bf{v}-\bf{u}$ has no physical 
meaning becomes evident by examining the following example. Let the velocities $\bf{u}$ and $\bf{v}$ 
have opposite directions and have magnitudes near to the speed of light or equal to it. The  
`velocity' $\bf{w}$ will have a magnitude near or equal to twice the speed of light, which is 
evidently absurd.
\end{quotation}
In special relativity every physical velocity must satisfy
\begin{flalign}
v^2_1 +v^2_2 +v^2_3 < 1
\label{vincolo_c}
\end{flalign}
in every inertial frame. Clearly, massless particles 
propagate at the velocity of
light, thus the relative velocity of two photons, or an electron and a photon is equal to 1 in 
every inertial frame.

Strictly speaking, following Fock, the mathematical expressions $|\boldsymbol{v}_1 
-\boldsymbol{v}_2|$ and $\bar{v}$ are not even  velocities in special relativity.

\subsection{Transformation of densities}

The product $n_1 n_2$ is not Lorentz invariant. In fact
the densities in a generic 
frame do not coincide with the proper densities because of the volume contraction, or, from another 
point of view, 
densities are only the time component of a 4-vector. 
Nonetheless, the product $n_1 n_2 \bar{v}$ is the physical invariant flux.
This means that 
the true relative velocity is hidden in this expression and  some cancellation takes place.
We shall see that this is what happens.

\subsection{The Landau-Lifschitz solution}

An answer to the above problems is given in \cite{Landau:1982dva}.
Let us rephrase the somewhat cumbersome  Landau--Lifschitz reasoning. 

They assume that both the cross section and the relative velocity  
must be defined in the rest frame of one particle, say particle 1 to fix the ideas.
In this frame, by definition of  rate, we have
$\mathcal{R}=\sigma v_\text{rel} n_1 n_2 =\sigma v_2 n^0_1 n_2$. 
In another frame, in general, the rate takes the form $\mathcal{R}=A n_1 n_2$, 
with $A$ a factor reducing to $\sigma v_\text{rel}$ in the rest frames.
We have to find the expression of $A$ in a generic frame.

Since the rate is invariant, the product $An_1 n_2$ must be invariant.
Using $n=\gamma n^0$, with $n^0$ the proper density corresponding to the rest frames, 
and $\gamma=E/m$, we have  
\[
A n_1 n_2=A\frac{E_1}{m_1}\frac{E_2}{m_2}n^0_1 n^0_2.
\]
The masses and the proper densities $n_0$ are numbers, thus
we must require that $A E_1 E_2=c_i$, with $c_i$ an invariant. 
Divide now both sides of the previous equality by 
the scalar $p_1\cdot p_2$, thus 
\[
A \frac{E_1 E_2}{p_1\cdot p_2}=c'_i,
\] 
with $c'_i =c_i/p_1\cdot p_2$  another invariant. 
In the rest frame of one particle
$E_1 E_2/p_1\cdot p_2=1$ and  $A=c'_i=\sigma v_\text{rel}$ by assumption. 
Hence in a generic frame 
\begin{flalign}
A=\frac{p_1\cdot p_2}{E_1 E_2} \sigma v_\text{rel}.
\label{Afactor}
\end{flalign}
In the rest frame of one particle 1 we already said that  $v_\text{rel}=|\boldsymbol{v}_2|$.
The 4-momenta are $p_1 =(m_1, 0)$, $p_2=(E_2,\boldsymbol{p}_2)$ with scalar product $p_1 
\cdot p_2 =m_1 E_2$. It follows that
$v_\text{rel} 
={|\boldsymbol{p}_2|}/{E_2}={\sqrt{E^2_2 -m^2_2}}/{E_2}$ can be written as\footnote{Formula (\ref{vrel_rest_frame_inv}) was also given in \cite{hagedorn1973relativistic}.}
\begin{flalign}
v_\text{rel} 
=\frac{\sqrt{(p_1 \cdot p_2)^2 -m^2_1 m^2_2}}{p_1\cdot p_2}.
\label{vrel_rest_frame_inv}
\end{flalign}
Using (\ref{v_moller_v}) and $p_1\cdot p_2/E_1 
E_2=1-\boldsymbol{v}_1 \cdot \boldsymbol{v}_2$, in terms of velocities
Eq.~(\ref{vrel_rest_frame_inv}) 
becomes
\begin{flalign}
v_{\text{rel}}=
\frac{
\sqrt{(\boldsymbol{v}_1 - \boldsymbol{v}_2)^2 - 
(\boldsymbol{v}_1 \times \boldsymbol{v}_2)^2}
}
{1-\boldsymbol{v}_1 \cdot \boldsymbol{v}_2}.
\label{v_rel_Rel_def}
\end{flalign} 
From Eq.~(\ref{Afactor}) and (\ref{vrel_rest_frame_inv}) the invariant flux  is 
\begin{flalign}
F=n_1 n_2 \frac{\cancel{p_1\cdot p_2}}{E_1 E_2} \frac{\sqrt{(p_1 \cdot p_2)^2 -m^2_1 m^2_2}}
{\cancel{p_1\cdot p_2}},
\nonumber
\end{flalign}
which gives Eq.~(\ref{flux_standard_n1n2}), or, in terms of velocities,
\begin{flalign}
F =n_1 n_2 \cancel{(1-\boldsymbol{v}_1 \cdot \boldsymbol{v}_2)}\frac{
\sqrt{(\boldsymbol{v}_1 - \boldsymbol{v}_2)^2 - 
{(\boldsymbol{v}_1 \times \boldsymbol{v}_2)^2}}}
{\cancel{1-{\boldsymbol{v}_1 \cdot \boldsymbol{v}_2}}},
\nonumber
\end{flalign}
which coincides with Eq.~(\ref{flux_standard_n1n2_Moller}).
Landau-Lifshitz\footnote{They
attribute  Eq.~(\ref{flux_standard_n1n2_Moller}) to Pauli in 1933, 
well before the appearance M\o{}ller's paper in 1945.
They do not give any specific reference and we could not find any Pauli's paper
where such formula is written.}  thus provide an \textit{ab initio} derivation
of the M\o{}ller formula (\ref{flux_standard_n1n2_Moller}).
A similar discussion was also given in~\cite{Terrall}.

We have highlighted the cancellation because is the central point.
Formula (\ref{flux_standard_n1n2_Moller}), and its particular case (\ref{flux_books}),
arises because there is the cancellation between 
the factor ($1-\boldsymbol{v}_1 \cdot \boldsymbol{v}_2$) that comes from the transformation of 
densities with the same factor  in the denominator of the relative velocity.
Only \textit{a posteriori}
we can say that the relativistic flux with collinear velocities is given by Eq.~(\ref{flux_books}).
The result of the cancellation is an expression which formally looks the nonrelativistic flux, 
but the relative velocity in the collinear case is 
\begin{flalign}
v^\parallel_{\text{rel}}=
\frac{
|\boldsymbol{v}_1 - \boldsymbol{v}_2| }
{1-\boldsymbol{v}_1 \cdot \boldsymbol{v}_2},
\label{v_rel_collinear}
\end{flalign}
not $|\boldsymbol{v}_1 - \boldsymbol{v}_2|$ as it is generally believed.

What about the `M\o{}ller velocity'? It is just 
the numerator of formula (\ref{v_rel_Rel_def}) or the product of the factor $(1-\boldsymbol{v}_1 
\cdot \boldsymbol{v}_2)$ with $v_\text{rel}$. This explains why 
it does not have  any physical meaning by itself. 
In the next sections we shall see that leaving such factor and the relative velocity 
explicitly in the formulas  allows for a clearer understanding of relativistic 
physics.
 
Actually it is possible to give a much general and simpler formulation than 
the Landau-Lifschitz's one
without any assumption about reference frames. But before, there are many interesting
properties of the relative velocity that is necessary to recall.

\section{Properties of the invariant relative velocity}
\label{sec:v_rel}

If we restore $c$ in (\ref{v_rel_Rel_def}), 
the vector product in the numerator and the scalar product in the denominator  are 
divided by $c^2$. In the nonrelativistic limit $c\to \infty$, $|\boldsymbol{v}_i|\ll c$
they disappear, and  $v_\text{rel}$ reduces to $v_r$. 
When one (or both) velocity is $c$, then $v_\text{rel} =c$, while  when both are smaller than 
$c$ then $v_\text{rel} < c$.
All the physical requirements are satisfied.

Formula (\ref{v_rel_Rel_def})  can be found in some books 
on relativity but to our knowledge
cannot be found in any particle physics or quantum field theory book.
Up to a minus sign in the numerator and in the denominator, 
this is the well known Einstein's rule for the composition 
of velocities already written in the 1905's paper \cite{Einstein:1905ve}. 

If one prefers to reason in terms of Lorentz transformations, 
the easiest way is to take $S$ as the laboratory frame
where $\boldsymbol{v}_1$ and $\boldsymbol{v}_2$ are given, and $S_1$ the rest frame of particle 
1. $S_1$ moves with velocity $\boldsymbol{v}_1$ with respect to $S$, hence the velocity 
$\boldsymbol{v}_{21}$ 
of particle 2 in $S_1$, that is the relative velocity, is found by applying a  boost 
$B(\boldsymbol{v}_1)$ to $\boldsymbol{v}_2$. 
One gets, see for example \cite{fock1969theory},~\cite{tsamparlis2010special},
\begin{flalign}
\boldsymbol{v}_{21}=
\frac{\boldsymbol{v}_2 -\boldsymbol{v}_1-(\gamma_1 -1)(1-\frac{\boldsymbol{v}_1\cdot\boldsymbol{v}_2}{\boldsymbol{v}^2_1})\boldsymbol{v}_1}
{\gamma_1(1-\boldsymbol{v}_1\cdot\boldsymbol{v}_2)}.
\end{flalign} 
By taking $S_2$ as the rest frame of particle 2, with similar reasoning,  instead we have
\begin{flalign}
\boldsymbol{v}_{12}=
\frac{\boldsymbol{v}_1 -\boldsymbol{v}_2-(\gamma_2 -1)(1-\frac{\boldsymbol{v}_1\cdot\boldsymbol{v}_2}{\boldsymbol{v}^2_2})\boldsymbol{v}_2}
{\gamma_2(1-\boldsymbol{v}_1\cdot\boldsymbol{v}_2)}.
\end{flalign} 
Differently from the nonrelativistic case where $\boldsymbol{v}_{12}=-\boldsymbol{v}_{21}$,
the vectors $\boldsymbol{v}_{12}$ and $\boldsymbol{v}_{21}$ belong to different directions
that differ by a spatial rotation.\footnote{ 
This is the Thomas-Wigner rotation 
and  corresponds to the well known fact that the product of  two non collinear boosts
gives a boost times a spatial rotation. See for example \cite{tsamparlis2010special}, 
\cite{gourgoulhon2013special}.
}
What is important for scattering theory  is that the magnitude of the two vectors is the 
same and symmetrical in indices 1 and 2, being equal to the relative velocity (\ref{v_rel_Rel_def})
\begin{flalign}
|\boldsymbol{v}_{12}|=|\boldsymbol{v}_{21}|=v_\text{rel},
\end{flalign}
as can be verified with direct calculation.

\subsection{Metric and hyperbolic properties}

In nonrelativistic physics 
the vectors $\bm{v}=(v_1,v_2,v_3)$ are points of the Euclidean  3-dimensional velocity space 
$\mathcal{V}_{\text{nr}}$.
The relative velocity ${v}_{\textit{r}} = \rvert \boldsymbol{v}_1 - \boldsymbol{v}_2 \rvert$
is invariant under Galileo transformations and 
 coincides with the Euclidean distance between the two 
points of $\mathcal{V}_{\text{nr}}$,
\begin{flalign}
v_r =\rho_{E}.
\label{vr_distance}
\end{flalign}

In special relativity  the velocity space $ \mathcal{V}_{SR}$ is subject to the constraint 
(\ref{vincolo_c}).
The space $\mathcal{V}_{SR}$ is given by the  points in the interior of sphere of unit radius, in 
natural units.
As it is well known, this is a Lobachevsky-Bolyai hyperbolic space with constant negative 
curvature \cite{fock1969theory}, 
\cite{Landau:1982dva}.

Using Eq. (\ref{v_rel_Rel_def}) we can calculate the relative velocity between two infinitesimally 
near points 
$\boldsymbol{v}$ and $\boldsymbol{v}+d\boldsymbol{v}$. This corresponds to the Riemann metric
\begin{flalign}
d\rho^2_H=\frac{(d\boldsymbol{v})^2-(\boldsymbol{v}\times d\boldsymbol{v})^2}{(1-\boldsymbol{v}^2)^2},
\label{riemmann}
\end{flalign}
where the subscript $H$ stands for hyperbolic.
As shown in \cite{fock1969theory}, the geodesics of $\mathcal{V}_{SR}$ 
are straight lines and 
the points of the segment between  $\boldsymbol{v}_1$ and $\boldsymbol{v}_2$ along the geodesic
can be parametrized by linear relations as 
\begin{flalign}
\boldsymbol{v}_\lambda=\boldsymbol{v}_1 + \lambda (\boldsymbol{v}_2 -\boldsymbol{v}_1),
\label{segment}
\end{flalign}
with $\lambda$  a continuous parameter varying in the interval $[0,1]$. 
Using (\ref{segment}) in (\ref{riemmann}) we find the line element 
\begin{flalign}
d\rho_H=\frac{\bar{v}\, d\lambda}{1-\boldsymbol{v}^2_\lambda}.
\end{flalign}
The length of the segment gives the distance between the two points of $\mathcal{V}_{SR}$.
Performing the integration we find
\begin{flalign}
\rho_H=\int^1_0 \frac{\bar{v}\, d\lambda}{1-\boldsymbol{v}^2_\lambda}=\frac{1}{2}\ln\left( \frac{1+v_\text{rel}}{1-v_\text{rel}}  \right).
\label{distance_VSR}
\end{flalign}
This is equal to $\tanh^{-1} ({v_\text{rel}})$; 
hence, the relation between the relative velocity and the distance is 
\begin{flalign}
v_\text{rel}=\tanh \rho_H,
\label{vrel_distance}
\end{flalign}
which represents the relativistic analogous of (\ref{vr_distance}).

Every velocity in $\mathcal{V}_{SR}$
can be thought as a relative velocity with respect to the origin $O$ 
with magnitude given by the distance from the origin.
In physics this distance  is called rapidity, and is indicated 
commonly with  $y$ or $\eta$. From (\ref{vrel_distance}) we have the usual relations
\begin{flalign}
v=\tanh y,\;\; \gamma=\cosh y,\;\; v \gamma =\sinh y.
\end{flalign}
We now eliminate the vector product in (\ref{v_rel_Rel_def}) using the identity 
(\ref{vector_identity}), 
\begin{flalign}
v_{\text{rel}}= 
\sqrt{1- 
\frac{(1-\boldsymbol{v}_1^2)(1-\boldsymbol{v}^2_2)}
{(1-\boldsymbol{v}_1 \cdot \boldsymbol{v}_2)^2}},
\label{v_rel_Rel_def_2}
\end{flalign}
and associate the  Lorentz factor to $v_\text{rel}$,
\begin{flalign}
\gamma_{{\text{r}}}=\frac{1}{\sqrt{1-v^2_{\text{rel}}}}.
\label{lorentz_factor_r}
\end{flalign}
From (\ref{v_rel_Rel_def_2}) and (\ref{lorentz_factor_r}) it follows the
fundamental relation
\begin{flalign}
\gamma_{\text{r}}&=\gamma_1 \gamma_2(1-\boldsymbol{v}_1 \cdot \boldsymbol{v}_2).
\label{Lorentz_factor_velocities_rep}
\end{flalign}
If $\theta$ is the angle between $\boldsymbol{v}_1$ and $\boldsymbol{v}_2$,
then Eq.~(\ref{Lorentz_factor_velocities_rep}) can  be written as
\begin{flalign}
\cosh y_\text{r} =\cosh y_1  \cosh y_2 -\sinh y_1 \sinh y_2 \cos\theta,
\label{cos_rule}
\end{flalign}
which is the cosine rule  of hyperbolic geometry.
When the velocities are collinear, this give the well known fact that rapidities sum 
up $y=y_1 \mp y_2$.

\begin{figure}
\includegraphics*[scale=0.45]{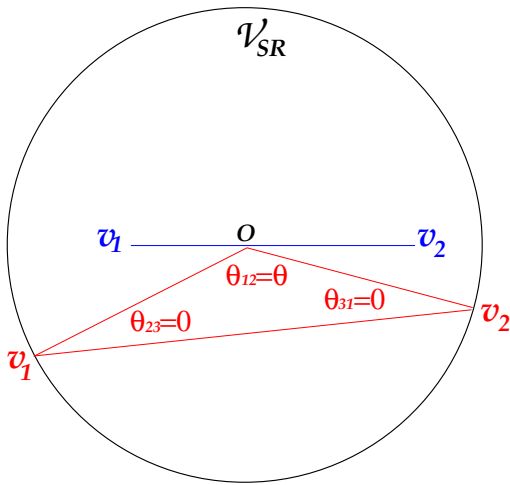}
\caption{The hyperbolic plane of velocity space represented by the Beltrami-Klein 
interior disk model of radius 1. 
The blue segment is relative to two massive particles 
colliding collinearly.
The red triangle  describes the scattering of two
ultrarelativistic particles colliding with angle $\theta$ in the laboratory.
}
\label{Fig:velocity_triangles}
\end{figure}

In Figure \ref{Fig:velocity_triangles} the blue segment passing for the origin describes the 
scattering of two massive particles 
with  collinear velocities with magnitudes $\boldsymbol{v}_1$ and $\boldsymbol{v}_2$.  
Using (\ref{distance_VSR}), the distance between the velocities 1 and 2 is the sum 
of the lengths $\rho_{1O}$ and $\rho_{2O}$
\begin{flalign}
\frac{1}{2}\ln\left( \frac{1+v_\text{rel}}{1-v_\text{rel}}  \right)=
\frac{1}{2}\ln\left( \frac{1+v_1}{1-v_1}  \right)+
\frac{1}{2}\ln\left( \frac{1+v_2}{1-v_2}  \right),
\nonumber
\end{flalign}
which gives  
\begin{align}
v_\text{rel}=\frac{v_1 + v_2}{1+v_1 v_2}.
\end{align}
The same result is obtained by (\ref{v_rel_collinear}) orienting the collision axis along
$\boldsymbol{v}_1$ and taking $\boldsymbol{v}_2$ in the opposite direction.

The metric (\ref{riemmann}) also determines the angles in velocity space \cite{fock1969theory}, 
\begin{flalign}
&\cos \theta_{ij}=\nonumber\\
&\frac{(\boldsymbol{v}_i-\boldsymbol{v}_k)\cdot(\boldsymbol{v}_j -\boldsymbol{v}_k)
-(\boldsymbol{v}_i\times \boldsymbol{v}_k)\cdot(\boldsymbol{v}_j\times\boldsymbol{v}_k) }
{\sqrt{(\boldsymbol{v}_i - \boldsymbol{v}_k)^2 - 
(\boldsymbol{v}_i \times \boldsymbol{v}_k)^2}
\sqrt{(\boldsymbol{v}_j - \boldsymbol{v}_k)^2 - 
(\boldsymbol{v}_j \times \boldsymbol{v}_k)^2}
},
\label{angles}
\end{flalign}
where angles are the vertices of the velocity triangle individuated by the points 
$\boldsymbol{v}_1$, $\boldsymbol{v}_2$ and $\boldsymbol{v}_3$, which must be taken in 
ciclic permutation in Eq.~(\ref{angles}).

In Figure \ref{Fig:velocity_triangles} we consider the example of scattering of ultrarelativistic 
or massless particles 
with velocities $\boldsymbol{v}_1=\hat{v}_1$ and $\boldsymbol{v}_2=\hat{v}_2$ 
in the laboratory frame at rest identified by the origin with velocity  $\boldsymbol{v}_3=0$.
Two  vertices of the red triangle are at infinity and the sides 
are three relative velocities equal to 1, that is of 
infinite length. From formula (\ref{angles}) 
it is easy to see that the two angles at the ideal vertices, $\theta_{23}$ and $\theta_{31}$, 
are zero,\footnote{The interior of the unit ball is the 3-dimensional extension of the Beltrami-Klein 
interior disk model for the hyperbolic plane
where the distance (\ref{distance_VSR}) corresponds to the Cayley-Klein projective distance, see 
for example \cite{Barrett:2011}.
The metric (\ref{riemmann}) is non conformal, thus the hyperbolic angles given by (\ref{angles}) in 
general do not coincide with the Euclidean angles between two velocities.
Only when one of the velocities corresponds with the origin, the angle at that vertex is equal to 
the Euclidean one. For considerations on the velocity space employing the conformal 
Beltrami-Poincar\'{e}
model see for example \cite{Rhodes:2004}.
}
while the angle at the origin $\theta_{12}$ is equal to the angle 
$\cos\theta=\hat{v}_1\cdot \hat{v}_2$ individuated by the velocities in the 
laboratory. In hyperbolic geometry the angles determine the sides of the triangle 
and their sum is less than $\pi$ by an amount given by the hyperbolic defect 
$\delta=\pi-(\theta_{12} +\theta_{13}+ \theta_{23})$. The defect gives the area of the 
triangle $A=K^2\delta$ where $K=1/\sqrt{-k}=1$ being $k=-1$ the gaussian curvature.\footnote{The 
Thomas-Wigner angle is related to 
the defect  and the area, see for example 
\cite{Rhodes:2004}, \cite{Barrett:2011}.}
The area of the red triangle in Figure \ref{Fig:velocity_triangles} is thus given by $A=\delta=\pi 
-\theta$.

\subsection{Manifestly invariant representations  }

Since $v_\text{rel}$ is Lorentz invariant it can be written
in terms of scalar products of various 4-vectors. 

In terms of the 4-velocity $u=\gamma(1,\boldsymbol{v})$, $u^2=1$, we have
\begin{flalign}
v_{\text{rel}}&=\frac{\sqrt{(u_1\cdot u_2)^2-1}}{u_1\cdot u_2}.
\label{v_rel_u}
\end{flalign}
The 4-momentum representation 
is given by (\ref{vrel_rest_frame_inv}), 
and in terms of the 4-current (\ref{4-current}) we obtain 
\begin{flalign}
v_{\text{rel}}&=\frac{\sqrt{(J_1 \cdot J_2)^2 - (J_1)^2 (J_2)^2}}{J_1 \cdot J_2}.
\label{v_rel_J}
\end{flalign}
Another useful formula is found introducing the Mandelstam variable  $s=(p_1+p_2)^2$,
\begin{flalign}
v_{\text{rel}} 
=\frac{\sqrt{\lambda(s,m^2_1,m^2_2)}}    {s-(m^2_1+m^2_2)},
\label{vrel_s}
\end{flalign}
where we used the triangular function 
\begin{flalign}
\lambda(s,m^2_1,m^2_2)=[s-(m_1 + m_2)^2][s-(m_1 - m_2)^2]. 
\label{Mandelstam_l}
\end{flalign}
For example the Mandelstam representation gives $s$ as a function of ${v}_\text{rel}$  
through the Lorentz factor $\gamma_\text{r}$ 
\begin{flalign}
s=(m_1 - m_2)^2 +2 m_1 m_2 (1+\gamma_{\text{r}}),
\label{s_gamma}
\end{flalign}
which is useful for cross section calculations.
Also the Lorentz factor can be written in terms of invariants as
\begin{flalign}
\gamma_{\text{r}}&=u_1 \cdot u_2=
\frac{p_1 \cdot p_2}{m_1 m_2}=\frac{J_1 \cdot J_2}{n^0_1 n^0_2}=
\frac{s-(m^2_1+m^2_2)}{2 m_1 m_2}.
\label{gamma_r_s}
\end{flalign}

Concluding this section we want to emphasize that the connection
with the metric properties of the velocity space given by Eq.~(\ref{vr_distance})
and Eq.~(\ref{distance_VSR})
shows that the relative velocity is a concept that is a logical consequence of the  
relativity principle.
The distance, hence the relative velocity, is a number that does not depend on the coordinate system 
or reference frame, both in the nonrelativistic and in the relativistic case.

\section{Lorentz invariant definition of flux }
\label{sec:flux}

Let us now go back to the incident flux and present how the Landau--Lifschitz reasoning
can be made simpler.

The nonrelativistic expression (\ref{flux_nonrel})  can only by used as a limit to which the new 
expression reduces in the nonrelativistic limit. 
The product $n_1 n_2$ must be replaced by some Lorentz scalar that reduces 
to $n_1 n_2=n^0_1 n^0_2 $.
The number densities are the time component of the 4-current (\ref{4-current}),
hence we are led to take the scalar product 
\[
J_1 \cdot J_2=n_1 n_2 (1-\boldsymbol{v}_1 \cdot \boldsymbol{v}_2)=
\frac{n_1}{\gamma_1}\frac{n_2}{\gamma_2}\gamma_\text{r}
=n^0_1 n^0_2 \gamma_\text{r},
\]
where we used Eq.~(\ref{Lorentz_factor_velocities_rep}).
This is the only scalar that can be formed with two 4-currents and presents the correct 
nonrelativistic limit.
Obviously $v_\text{rel}$ takes the place of $v_r$.
It follows that the natural definition of the relativistic invariant flux is 
\begin{flalign}
F=(J_1 \cdot J_2)v_{\text{rel}}.
\label{rel_flux}
\end{flalign}
This expression is a Lorentz scalar at sight and does not rely on the rest frame or the center of 
momentum frame.\footnote{A similar discussion was given in \cite{Weaver} (where anyway the 
rest frame is used), \cite{Furman} and \cite{Cannoni:2015wba}.}
For  massless particles the velocity vector becomes the unitary vector in the direction of 
propagation and when at least one massless particle is involved in the scattering then
$v_\text{rel}=1$.
For  two massless particles, say two photons,
the flux reads $F=n_1 n_2 (1-\cos\theta)$,
with $\theta$ the angle between $\hat{k}_1$ and $\hat{k}_2$.
For collisions of a massless with a massive particle, $F=n_1 n_2 (1-v_2\cos\theta)$. 

When computing cross sections in quantum field theory, the 4-momentum representation 
(\ref{vrel_rest_frame_inv})
is more useful. We can write
\begin{flalign}
F=4(p_1 \cdot p_2) v_\text{rel},
\end{flalign}
and when at least one massless particle is involved $F=4(p_1 \cdot p_2)$.
Clearly expression (\ref{flux_standard}) is only one of the many ways
the invariant flux (\ref{rel_flux}) can be explicitly written. For example, 
using the Mandelstam variable representation one easily finds the well known expression
$F=2\sqrt{\lambda(s,m^2_1,m^2_2)}$.

While formula (\ref{rel_flux}) explicitly displays the invariance property,
in order to highlight the physical content it is more useful write
\begin{flalign}
F=n_1 n_2 k_\text{r}v_{\text{rel}},
\label{flux_mydefinition}
\end{flalign}
where we have defined the \textit{hyperbolic correlation factor}
\begin{flalign}
k_\text{r}= 1-\boldsymbol{v}_1 \cdot \boldsymbol{v}_2 
=\frac{\gamma_\text{r}}{\gamma_1 \gamma_2}
=\frac{\cosh y_\text{r}}{\cosh y_1 \cosh y_2}
=\frac{p_1 \cdot p_2}{E_1 E_2}.
\label{k_r}
\end{flalign}
The relative velocity (\ref{v_rel_Rel_def}) hence enters two times in the incident flux: 
explicitly as a factor and implicitly through $k_\text{r}$ that is nothing but the hyperbolic 
cosine rule given by Eqs. (\ref{Lorentz_factor_velocities_rep}) and (\ref{cos_rule}).
The hyperbolic correlation factor $k_\text{r}$ is not a scalar, only the product $n_1 n_2 k_\text{r}$ 
is Lorentz invariant, and attains the maximum value of 2 for example in  the case of head on 
collinear scattering of two photons.

\section{Flux and luminosity at colliders}
\label{sec:luminosity}

We now discuss how experimentalists in high energy particle physics use the concepts of 
invariant cross section and flux.

In collider physics it is common to consider the rate integrated over the interaction volume.
Using the expression Eq.~(\ref{flux_mydefinition}) for the flux we have
\begin{flalign}
\frac{dN_f}{dt} 
=\sigma \int dV n_1 n_2 k_\text{r} v_\text{rel}=\sigma\mathcal{L}(t),
\label{lumi_def}
\end{flalign}  
which defines the instantaneous luminosity $\mathcal{L}(t)$.
The integrated luminosity $\mathcal{L}_\text{int} =\int dt\mathcal{L}(t)$ gives the total number 
of expected events $N_f=\sigma \mathcal{L}_\text{int}$ in a certain running time. 

\begin{figure}
\includegraphics*[scale=0.38]{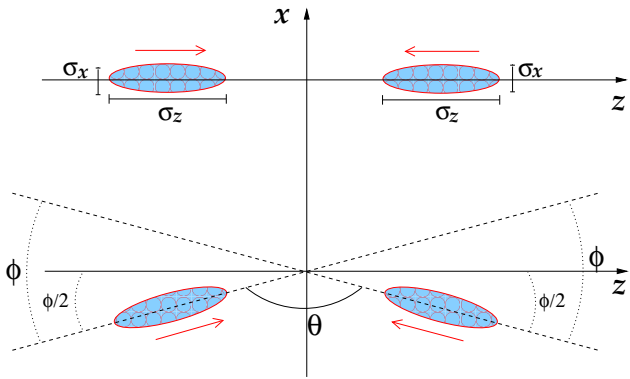}
\caption{Schematic representation of two bunches in a collinear head-on collision,
top figure, and collision with a crossing angle $\phi$, bottom figure.}
\label{Fig:crossing}
\end{figure}

In  storage rings like the Large Hadron Collider (LHC)
the beams are constituted by  bunches with 
$\texttt{n}$  particles and  Gaussian shape 
characterized by spatial dispersions $\sigma_{x},\sigma_{y},\sigma_{z}$ 
(root mean squared deviations) along the three directions. 
Assume further that the beams move along the $z$ axis in opposite 
directions with ultrarelativistic velocity $v\sim 1$ 
and produce head-on collisions at the origin of the $z=0$ as in Figure \ref{Fig:crossing}.
The number densities are thus given by
\begin{flalign}
n_\pm (\boldsymbol{x},t)=\frac{\texttt{n}}{(2\pi)^{3/2}\sigma_x \sigma_y \sigma_z}
e^{-(\frac{x^2}{2\sigma^2_x}+\frac{y^2}{2\sigma^2_y}+\frac{(z\pm vt)^2}{2\sigma^2_z})}.
\label{gaussian}
\end{flalign} 
The luminosity is then
\begin{flalign}
\mathcal{L}_\parallel(t)=f N_b \int d^3 \boldsymbol{x}\, n_+(\boldsymbol{x},t) n_- (\boldsymbol{x},t)
k_\text{r} v_\text{rel}, 
\label{L_int}
\end{flalign}
where we have multiplied by the revolution frequency $f$ and the number of bunches $N_b$. 
The parallel symbol indicate the collinear beams.
Performing the Gaussian integrations in $x$, $y$, $s=vt$,
we obtain  the well 
known result also reported by the Particle Data Group~\cite{Agashe:2014kda}
\begin{flalign}
\mathcal{L}_\parallel (t)=2 \frac{\texttt{n}_1 \texttt{n}_2 f N_b}{8\pi \sigma_x \sigma_y} .
\label{luminosity_collinear}
\end{flalign}
The factor 2 arises because for ultrarelativistic particles $v\simeq 1$,
$v_\text{rel}\simeq 1$ and the velocities correlation factor for this configuration 
is $k_\text{r}\simeq 2$, thus\footnote{
We think that this way of understanding the factor 2 is much more clear and physical 
than saying that the 
M\o{}ller factor is 2. Note that collider physicists
do not use the word `velocity' for the M\o{}ller expression but 
more correctly speak of kinematic or luminosity factor, see for example
 \cite{Middelkoop:1963zz}, \cite{Napoly:1992kn}, \cite{Furman}, \cite{Herr:2003em}.
}
$k_\text{r} v_\text{rel}\simeq 2$. 

In reality at the LHC beams are not collinear but there is an angle $\theta$  between them.  
Collider physicist prefer to use the complementary angle $\phi=\pi-\theta$, called crossing angle,
see Figure~\ref{Fig:crossing}.
The crossing angle is necessary to confine the interaction region and to avoid unwanted collisions
and reduce other effects, see for example \cite{Herr:2003em}. 
While at LHC the crossing angle is small, around 300 $\mu$rad,
at the old Intersecting Storage Ring at CERN, where beams were unbunched, it was about 18$^\circ$.

The hyperbolic correlation factor with crossing angle and ultrarelativistic velocities reads
\begin{flalign}
k_\text{r}\simeq 1-\cos\theta =1+\cos\phi =2 \cos^2\frac{\phi}{2}.
\end{flalign}
If the beams are contained in the $x,z$ plane with $\sigma_z \gg\sigma_x,\sigma_y$, see Figure~\ref{Fig:crossing},
the calculation gives \cite{Herr:2003em}
\begin{flalign}
\mathcal{L} (t,\phi)=\mathcal{L}_\parallel (t) S(\phi) ,
\label{Lumi_phi}
\end{flalign}
where
\begin{flalign}
S(\phi)= \frac{1}{\sqrt{1+(\frac{\sigma_z}{\sigma_x}\tan \frac{\phi}{2})^2}}
\end{flalign}
is the reduction factor. 
For example, at the LHC, the crossing angle is $\phi=285$ $\mu$rad
and $\sigma_z=7.7$ cm, $\sigma_x \simeq 16.7$ $\mu$m, that give $S=0.835$~\cite{Herr:2003em}.
Note that the crossing angle $\phi$ in the laboratory is equal to the hyperbolic defect 
of the velocity triangle in velocity space of Figure~\ref{Fig:velocity_triangles}.

At this point it is necessary to make two remarks.

The quantity 
$A=4\pi \sigma_x \sigma_y  $ that appears in Eq.~(\ref{luminosity_collinear}), or $A/S(\phi)$ in 
Eq.~(\ref{Lumi_phi}),
can be considered as an effective  cross sectional area of the bunches
transverse to the plane that contains the beams, 
but this quantity depends on the details of the machine and  has nothing to do with the 
theoretical cross section as we argued in footnote 3.

The crossing angle affects the luminosity $\mathcal{L}_\phi$ and 
the number of observed events $\dot{N}_f$ but not the ratio, 
that is the extracted cross section $\sigma_\text{exp}=\dot{N}_f / \mathcal{L}_	\phi$, 
which is an invariant quantity.
The theorist does not have to care about the fact that real beams have a crossing angle
if the cross section $\sigma^\text{th}=\mathcal{R}^\text{th}_f /F$ is invariant. For this reason
the flux must be invariant under general Lorentz transformations and not only under boosts
along a fixed direction. The invariant cross section can be calculated in any frame and 
compared with $\sigma_\text{exp}$.

\section{Relative velocity as kinematic variable }
\label{sec:kin_var}

The relative velocity $v_\text{rel}$ and the Lorentz factor $\gamma_\text{r}$ 
are good variables to express the invariant cross 
section as much as scalar product of 4-momenta,  Mandelstam variables or rapidity.
For a general example we 
consider the quantum electrodynamics processes $e^+ e^- \to \gamma \gamma$, pair annihilation,
and its inverse $\gamma \gamma \to e^+ e^- $, pair creation.

The total annihilation cross section, see for example~\cite{berestetskii2012quantum}, can be written 
as a function of the variable $\tau={s}/{4m^2}$ as
\begin{flalign}
\sigma_{\text{ann}}(\tau) 
=\frac{\sigma_0} {2\tau^2(\tau-1)}
&\left[
(\tau^2+\tau-\frac{1}{2})\ln\frac{\sqrt{\tau}+\sqrt{\tau-1}}{\sqrt{\tau}-\sqrt{\tau-1}}\right.\nonumber\\
&\left. -(\tau+1)\sqrt{\tau}\sqrt{\tau-1}
\right],
\end{flalign}
where $\sigma_0=\pi \alpha^2/m^2$, $\alpha$ the fine structure constant and $m$ the electron's mass. 
Using Eq.~(\ref{s_gamma}) with $m_1 =m_2 =m$ we obtain
\begin{flalign}
s=2m^2(1+\gamma_\text{r}),\;\;\;
\label{s_gamma_m1=m2}
\tau=\frac{1}{2}(1+\gamma_\text{r}),
\end{flalign}
and the cross section can be written as
\begin{flalign}
\sigma_{\text{ann}}(\gamma_\text{r}) 
=\frac{\sigma_0}{1+\gamma_\text{r} }
&\left[
\frac{\gamma^2_\text{r} +4\gamma_\text{r}+1}{\gamma^2_\text{r} -1}
\ln(\gamma_\text{r} +\sqrt{\gamma^2_\text{r}-1})
\right.\nonumber\\
&\left. 
-\frac{\gamma_\text{r} +3}{\sqrt{\gamma^2_\text{r}-1}}
\right].
\label{dirac_cross_section_gamma}
\end{flalign}
This expression is valid in any frame. 
In the rest frame of the electron to fix the ideas, the relative velocity coincides with
positrons velocity  $v_{+}$, hence
\begin{flalign}
v_\text{rel}(v_+)=v_+, 
\;\;\;\;\gamma_\text{r}(v_+)&=\gamma(v_+).
\label{gamma_r_lab}
\end{flalign}
The  cross section in the rest frame of the electron is thus given by the same Eq. 
(\ref{dirac_cross_section_gamma}), 
with $\gamma_\text{r}$ replaced by $\gamma(v_+)$.

In the center of momentum frame instead Eq. (\ref{v_rel_Rel_def}) gives
\begin{flalign}
v_\text{rel}(v_*)&=\frac{2v_*}{1+v^2_*},
\label{v_rel_*}
\end{flalign}
with Lorentz factor
\begin{flalign}
\gamma_\text{r}(v_*)&=\frac{1+v^2_*}{1-v^2_*}.
\label{gamma_*}
\end{flalign}
Substituting (\ref{gamma_*}) into Eq.~(\ref{dirac_cross_section_gamma}) 
we find 
\begin{flalign}
\sigma_{\text{ann}}(v_*)=
\sigma_0 \frac{1-v^2_*}{4v_*}  
\left[
\frac{3-v^4_*}{v_*}\ln\frac{1+v_*}{1-v_*}-2(2-v^2_*)
\right].
\label{sigma_ee_gg_cmf}
\end{flalign}
In this way we have obtained the  known formula~\cite{berestetskii2012quantum}
using only the relative velocity.

The cross section of the inverse process~\cite{berestetskii2012quantum} is related to 
Eq.~(\ref{sigma_ee_gg_cmf}) by
\begin{flalign}
\sigma_{\text{pair}}(v_*) =    2 v^2_* \sigma_{\text{ann}}(v_*).
\end{flalign}
In order to find the expression valid in any frame we invert Eq.~(\ref{gamma_*})
\begin{flalign}
v^2_*=\frac{\gamma_\text{r}-1}{\gamma_\text{r}+1}=\frac{\tau-1}{\tau}=1-\frac{2m^2}{k_1\cdot k_2},
\end{flalign}
where $k_i$ are the 4-momenta of the colliding photons.
For example, using $\tau$ as variable,  we can write
\begin{flalign}
\sigma_{\text{pair}}(\tau)=2
\frac{\tau-1}{\tau} \sigma_{\text{ann}}(\tau).
\end{flalign}

In the nonrelativistic limit $v_\text{rel}\sim v_r\ll 1$,
and the expansion of (\ref{dirac_cross_section_gamma}) in powers of the relative velocity 
reads
\begin{flalign}
\sigma^\text{nr}_{\text{ann}}(v_\text{rel})\sim 
\sigma_0 (\frac{1}{v_\text{rel}}-\frac{3}{20}v^3_\text{rel}-\frac{11}{105}v^5_\text{rel}+...).
\label{expansion_1}
\end{flalign}
The expansion in terms of $v_+$ in the rest frame of the electron has the same coefficients. 
Instead, the expansion of (\ref{sigma_ee_gg_cmf}) to the same order 
for small $v_*$ is 
\begin{flalign}
\sigma^{\text{nr}}_{\text{ann}}(v_*)\sim 
\sigma_0 (\frac{1}{2 v_*}+\frac{v_*}{2}-\frac{6}{ 5} v^3_*+ \frac{26}{105}v^5_*+...).
\label{expansion_2}
\end{flalign}
Both expansions follow the behavior $\sigma\propto\sum_\ell a_\ell v^{2\ell-1}$
as it is expected for inelastic exothermic processes at low energy.
At very low velocities the cross section is dominated by the $\ell=0$ partial wave or $S$-wave.
Anyway the coefficients are different, in particular in (\ref{expansion_1}) the one corresponding 
to $\ell=1$ is zero. 
Note that (\ref{v_rel_*}) has an expansion in odd powers of $v_*$ 
\begin{flalign}
v_\text{rel}=2v_*(1- v^2_* + v^4_* -v^6_* +...),
\label{sum_1}
\end{flalign}
while
\begin{flalign}
\frac{1}{v_\text{rel}}=\frac{1}{2v_*}+\frac{v_*}{2},
\label{sum_2}
\end{flalign}
has only two terms.
In order to obtain (\ref{expansion_2}) from (\ref{expansion_1}) it is necessary to substitute 
(\ref{sum_2}) in the first term of (\ref{expansion_1}) and to take powers of (\ref{sum_1}) 
keeping all the terms of the same order. 
For example, $v^3_\text{rel}\sim 8 v^3_* -24 v^5_*$ 
gives the term in $v^3_* $ and a contribution to the order $v^5_*$ that must be summed to
$v^5_\text{rel}\sim 32 v^5_*$. In this way the expansion (\ref{expansion_2}) is recovered.

If incorrectly we used the `M\o{}ller velocity' 
$\bar{v}_*=2 v_*$ in $\sigma^{\text{nr}}_\text{ann}(v_\text{rel})$, 
we would obtain wrong cross section and 
expansion in the center of mass frame. In terms of $s$ we have $\bar{v}_* =2 \sqrt{1 -4 m^2/s}$,
which gives the relation
\begin{flalign}
s =\frac{4 m^2}{1-\frac{\bar{v}^2_{*}}{4}}\sim 4m^2 + m^2 v^2_{r}+\frac{m^2}{4} v^4_{r}++\frac{m^2}{16} v^6_{r}+...,
\label{s_vr_cmf}
\end{flalign}
where the expansion corresponds to the nonrelativistic limit $\bar{v}_* \sim v_r\ll 1 $.
This is different from the expansion of (\ref{s_gamma_m1=m2}) 
\begin{flalign}
s \sim 4m^2+m^2 v^2_r +\frac{3}{4} m^2 v^4_r+\frac{5}{8} m^2 v^6_r+...,
\label{s_expansion_vrel}
\end{flalign}
which is the correct expansion of $s$ to use in the cross section to obtain
the nonrelativistic limit. 
The misuse of the  `M\o{}ller velocity'
thus can also be source of errors.
Examples in dark matter phenomenology are discussed in \cite{Cannoni:2015wba}.

\section{Relative velocity and  relativistic collisions in gases}
\label{sec:gas}

The hyperbolic correlation factor $k_\text{r}$ plays a fundamental role also in the 
theory of collisions in a relativistic gas.
We  consider  an ideal gas composed of two species of particles with mass $m_1$ and $m_2$
in thermal equilibrium at a given temperature $T$ and work in the frame
where the gas is at rest as a whole, the local rest frame or comoving frame. 
Before discussing the relativistic gas it is useful to briefly review 
the nonrelativistic case.

In kinetic theory the number density is determined by the Boltzmann one particle phase space 
distribution 
$f^\text{eq}=\exp(-E/T)$, $E=\boldsymbol{p}^2 /2m$, by the integral $n=g/(2\pi)^3 \int 
d^3\boldsymbol{p} f^\text{eq}$, 
\begin{flalign}
n^\text{eq}=g\left(\frac{mT}{2\pi}\right)^{3/2} e^{-m/T},
\label{n_Boltzmann}
\end{flalign}
with $g$ degrees of freedoms of the particle.

It is well known, see for example~\cite{landau2013statistical}, that the probability 
density function of the relative velocity $v_r$ such that $\int^{\infty}_{0} dv_r P(v_{{r}})=1$
is given by
\begin{flalign}
P(v_{{r}})=\sqrt{\frac{2}{\pi}}\left(\frac{\mu}{T}\right)^{3/2} 
{v}^2_{{r}}\, e^{-\frac{\mu}{T}\frac{v_{r}^2}{2}},
\label{Maxwell_vrel}
\end{flalign}
where $\mu =m_1 m_2/(m_1 +m_2)$ is the reduced mass. 
The mean value $\int^{\infty}_0 dv_r P (v_r) v_r$ of $v_r$ is given by
\begin{flalign}
\langle v_r \rangle_P=\sqrt{\frac{8 T}{\pi \mu}},
\label{vr_average_Maxwell}
\end{flalign}
and, being $\sigma_\text{nr}$ a nonrelativistic cross section that is function of $v_r$,
the averaged rate  is then 
\begin{flalign}
\langle \mathcal{R}\rangle= \frac{n^\text{eq}_1 n^\text{eq}_2} {j_{12}}
\langle \sigma_\text{nr} v_r\rangle_{P},\;\;j_{12}=1+\delta_{12},
\end{flalign}
where $\delta_{12}$ is the Kronecker's delta and the factor $j_{12}$ must be inserted to avoid 
double counting when the 1 and 2 are the same specie.
The so-called thermal averaged cross section is then
\begin{flalign}
\langle \sigma_\text{nr} v_r\rangle_{P}=
\sqrt{\frac{2}{\pi}}\left(\frac{\mu}{T}\right)^{3/2}
\int^\infty_0 dv_r {v}^2_{{r}}\, e^{-\frac{\mu}{T}\frac{v_{r}^2}{2}} \sigma_\text{nr} v_r.
\label{nr_average_def}
\end{flalign}
Formula (\ref{nr_average_def}) is commonly used for the calculation of 
abundances of dark matter particles that decoupled at a temperature
when they were nonrelativistic, see for example \cite{Griest:1990kh}.
Instead, expressed as a function of the relative kinetic energy
$E =1/2 \mu v^2_r$, Eq.~(\ref{nr_average_def}) takes the form
\begin{flalign}
\langle \sigma_\text{nr} v_r\rangle_{P}&=\sqrt{\frac{8}{\pi \mu T^3}}\int^\infty_0 dE \, E\, 
e^{-\frac{E}{T}}\, \sigma_\text{nr}(E) ,
\label{nr_average_energy}
\end{flalign}
which, for example, is used
for calculating  nuclear reaction rates in the Sun \cite{Adelberger:2010qa} 
and rates in big bang nucleosynthesis \cite{Iocco:2008va}.

\subsection{The relativistic classical gas}

There are other physical problems, for example the calculation of abundances of relics that 
decoupled when they were relativistic,  
the calculation of particle yields in ultrarelativistic ion collisions
and reaction rates in astrophysical plasmas 
where the nonrelativistic approximation is not good.

The relativistic generalization of the Boltzmann distribution, also known as J\"{u}ttner 
distribution~\cite{groot1980relativistic},~\cite{cercignani2002relativistic},
is
\begin{flalign}
f^\text{eq} = \exp({-{E}/{T}}),\;\;\;E=\sqrt{\boldsymbol{p}^2+m^2}.
\end{flalign}
The average number density  $n^\text{eq}=\frac{g}{(2\pi)^3}\int d^3\boldsymbol{p} f^\text{eq}$
in this case is given by 
\begin{flalign}
n^\text{eq}=\frac{g}{(2\pi)^3} 4\pi m^2 T K_2 (x),
\label{n_0}
\end{flalign}
where $x=m/T$ and $K_n(x)$ are modified Bessel functions of the second kind 
that  appear in almost all the formulas of relativistic statistical mechanics.\footnote{
The first terms of the asymptotic expansions for $x\gg 1$ and $x\ll1$ 
useful in the nonrelativistic and ultrarelativistic are respectively
\[
K_n (x)\sim e^{-x}\sqrt{\frac{\pi}{2x}}(1+\frac{4n^2-1}{8x}),\,\;\;\;
K_n (x)\sim \frac{(n-1)!}{2}\left(\frac{2}{x}\right)^n,
\]
with the latter valid for $n>0$.}

It is useful to introduce the normalized  momentum distribution 
\begin{flalign}
f_{p} (\boldsymbol{p})=\frac{1}{4\pi m^2 T K_2(x)}e^{-\sqrt{\boldsymbol{p}^2+m^2}/T},
\label{f_0,p}
\end{flalign}
such that $\int d^3 \boldsymbol{p}f_{p} (\boldsymbol{p})=1 $.
In the rest of the Section we will abbreviate the notation indicating 
$f_{p}(\boldsymbol{p}_i)\equiv f_{p,i}$. 
In this way the average value of a generic function of two momenta 
is given by $\int d^3\boldsymbol{p}_1 d^3\boldsymbol{p}_2 f_{p,1} f_{p,2} 
G(\boldsymbol{p}_1,\boldsymbol{p}_2) $.

The next step is to write the probability density function of $v_\text{rel}$, let us call it 
$\mathcal{P}(v_\text{rel})$.
It was shown in~\cite{Cannoni:2013bza}, and at this point it should not come as a surprise, 
that is the averaged value of the hyperbolic correlation factor 
\begin{flalign}
\langle k_\text{r} \rangle=\int {d^3 \boldsymbol{p}_1} {d^3 \boldsymbol{p}_2}
f_{p,1} f_{p,2}
\frac{p_1\cdot p_2}{E_1 E_2}
=\int^{1}_{0} dv_{\text{rel}} \mathcal{P}(v_{\text{rel}}) =1,
\label{mean_kr}
\end{flalign}
which determines the probability density function of $v_\text{rel}$,
\begin{flalign}
\mathcal{P}(v_{\text{rel}})&=
\frac{X
 \frac{\gamma^3_{_{\text{r}}} (\gamma^2_{_{\text{r}}} -1)}{\sqrt{\gamma_{{\text{r}}} +\varrho}}
K_1 (\sqrt{2}X\sqrt{\gamma_{{\text{r}}} +\varrho})
}
{\sqrt{2} \prod_{i} K_2 (x_i) },
\label{P_v}
\\
X&=\sqrt{x_1 x_2},\;\;\varrho=\frac{x^2_1 +x^2_2}{2x_1 x_2}.
\label{X_ro}
\end{flalign}

Thanks to (\ref{P_v}) the mean value $\int^{1}_{0} dv_{\text{rel}} 
\mathcal{P}(v_{\text{rel}})v_{\text{rel}} $
of the relative velocity is given by the expression \cite{Cannoni:2013bza}
\begin{flalign}
\langle v_{\text{rel}} \rangle_\mathcal{P}=
\frac{2}{\xi}
\frac{(1+\varrho)^2 K_3(\xi) - (\varrho^2-1) K_1 (\xi)}{ K_2 (x_1) K_2(x_2)},
\label{Vmean_x1_x2}
\end{flalign}
where $\xi=x_1 +x_2$. The ultra--relativistic limit $x\to 0$ of the mean value 
(\ref{Vmean_x1_x2}) is 1, and the fluctuations tend to zero, thus the bound imposed by the 
velocity of light is not violated even in the statistical sense.

The averaged relativistic rate is then obtained by integrating 
$\mathcal{R}=\frac{n_1 n_2}{j_{12}} k_\text{r} \sigma v_\text{rel}$  
over the momenta with distribution (\ref{f_0,p}). This equivalent
to averaging with the distribution of the relative velocity,
\begin{flalign}
\langle \mathcal{R} \rangle=\frac{n^\text{eq}_1 n^\text{eq}_2}{j_{12}}\langle k_\text{r} \sigma 
v_\text{rel}\rangle
= \frac{n^\text{eq}_1 n^\text{eq}_2}{j_{12}}\langle \sigma v_\text{rel}\rangle_\mathcal{P}.
\end{flalign}
The relativistic thermal averaged cross section is 
\begin{equation}
\langle \sigma_{} {v}_{\text{rel}} \rangle_\mathcal{P}
=\frac{
X \int^{1}_{0} dv_{\text{rel}} 
 \frac{\gamma^3_{_{\text{r}}} (\gamma^2_{_{\text{r}}} -1)}{\sqrt{\gamma_{{\text{r}}} +\varrho}}
K_1 (\sqrt{2}X\sqrt{\gamma_{{\text{r}}} +\varrho})\sigma v_{\text{rel}}
}
{\sqrt{2} \prod_{i} K_2 (x_i)}.
\label{sigmav_gamma}
\end{equation}
Changing variable from $v_\text{rel}$ to $s$ 
using equations (\ref{vrel_s}) and (\ref{s_gamma}), 
formula (\ref{sigmav_gamma}) 
takes the more popular form\footnote{
Formulas (\ref{sigmav_gamma})-(\ref{sigmav_s}) were found independently by physicist 
working in different areas.
Probably Eq.~(\ref{sigmav_gamma}), written in terms of the the variables  
$z=\sqrt{2}X\sqrt{\gamma_{_{\text{r}}} 
+\varrho}$ and $\chi=\gamma_{\text{r}} {v}_{\text{rel}}$,
was first obtained in \cite{Weaver} for applications to astrophysical 
relativistic plasmas. 
In the form (\ref{sigmav_s}) it appears in \cite{Biro:1981es} for applications in heavy-ion 
collisions physics, and in \cite{Claudson:1983js} for the study of a model for baryogenesis. Then 
(\ref{sigmav_s}) with $m_1=m_2=m$ was found again in \cite{Gondolo:1990dk} in studying the chemical 
decoupling of dark matter. Finally, remarking  the role of the invariant relative velocity and its 
probability distribution, was again obtained in \cite{Cannoni:2013bza}. 
}
\begin{flalign}
\langle \sigma_{} {v}_{\text{rel}} \rangle_\mathcal{P}
=\frac{\int^{\infty}_{(m_1+m_2)^2} ds 
\frac{\lambda(s,m^2_1,m^2_2)}{\sqrt{s}}
K_1 (\frac{\sqrt{s}}{T}) \sigma}{8 T \prod_i m^2_i K_2 (x_i)}.
\label{sigmav_s}
\end{flalign}
When  $x\gg 1$ all the relativistic formulas reduce
to the corresponding nonrelativistic expression.
The analogy between the nonrelativistic and the relativistic case is thus complete. 

\subsection{Thermal averaged cross section}

In the case of the Fermi-Dirac and Bose-Einstein distributions
it is not possible to obtain a close expression for the distribution of the relative velocity.
Quantum effects can be considered using the expansions
\begin{flalign}
f^{\text{F,B}}=\frac{1}{e^{E/T} \mp 1}=(\pm)\sum_{i=1}^{\infty}(\pm)^i e^{-i E/T},
\end{flalign}
which amounts to  the replacement
\begin{flalign}
K_1 \left(\frac{\sqrt{s}}{T}\right)\rightarrow 
\sum_{i=1}^{\infty}\sum_{j=1}^{\infty}\frac{(\pm)^{i+j}}{\sqrt{ij}}K_1 \left(\frac{\sqrt{ijs}}{T}\right),
\end{flalign}
in formula (\ref{sigmav_s}), see for example \cite{letessier2002hadrons}.

In general Eq.~(\ref{sigmav_s}) must be integrated numerically but
effective interactions of the type
\begin{flalign}
\mathcal{L}_\Lambda=\frac{\lambda_a \lambda_b}{\Lambda^2} 
(\bar{\chi} \Gamma_a \chi) (\bar{\psi}\Gamma_b  \psi),
\label{operator}
\end{flalign}
give cross sections that are simple enough and the analytical integration is possible 
\cite{Cannoni:2015wba}.
In Eq.~(\ref{operator}), $\lambda_{a,b}$ are dimensionless  coupling associated with the 
interactions described by a combination of Dirac matrices $\Gamma_{a,b}$.
$\Lambda$ is the energy scale below which the effective field theory is valid.
The $\chi$'s are Dirac or Majorana fields and
have mass $m$, while $\psi$ are fermions whose mass is much smaller than $m$ and $\Lambda$
such that can be considered massless with very good approximation, $m_\psi =0$. 
Under these assumptions, the cross section for 
$\chi\chi\to\psi\psi$ is given by
\begin{flalign}
\sigma=\sigma_\Lambda \frac{a}{2} \frac{\sqrt{s}}{\sqrt{s-4m^2}} \left(\frac{s}{4m^2} 
+\frac{k}{4}\right),\;\;\sigma_{{\Lambda}} =\frac{\lambda^2_a \lambda^2_b}{4\pi} 
\frac{m^2}{\Lambda^4},
\label{sigmatot_k}
\end{flalign}
while  $a$ and $k$ are numerical coefficients whose value  depends on the particular type of 
interactions and on the nature of the annihilating particles.

The thermal average (\ref{sigmav_s}) of  the cross section (\ref{sigmatot_k}) is given by 
\cite{Cannoni:2015wba}
\begin{flalign}
\langle \sigma {v}_{\text{rel}} \rangle_\mathcal{P}&=\sigma_\Lambda a \Phi_k(x),\label{sigmav_ann_simple}\\
\Phi_k (x)&=\frac{1 }{16}
\left(8 +2k + (5+2k)\frac{K^2_1(x)}{K^2_2(x)} + 3\frac{K^2_3(x)}{K^2_2(x)}\right).
\label{Sphi}
\end{flalign}
There are two particular cases that are worth to discuss in details for their frequent use
in dark matter model building and phenomenology.

If $k=0$, which is the case of $s$-channel
annihilation with  pseudoscalar interaction $\Gamma_a =\Gamma_b=\gamma^5$,  
we have
\begin{flalign}
\Phi_0 (x)&=
\frac{1}{16}
\left(8 + 5\frac{K^2_1(x)}{K^2_2(x)} + 3\frac{K^2_3(x)}{K^2_2(x)}\right)\\
&\sim 1 +\mathcal{O}(x^{-2}) \;\;x\gg 1,
\label{Pphi}
\end{flalign}
which describes $S$-wave scattering  nonrelativistic limit. 
In fact, the expansion of the cross section (\ref{sigmatot_k}) is $\sigma_\text{nr} v_r \sim 
\sigma_\Lambda a+O(v^4_r)$
thus the thermal average is constant and temperature independent in agreement with (\ref{Sphi}).

The value  $k=-4$ is found
in the case of $s$-channel annihilation 
with scalar $\Gamma_a =\Gamma_b=1$ and axial-vector $\Gamma_a =\Gamma_b=\gamma^\mu\gamma^5$ 
couplings, and $t$-channel Majorana fermion annihilation with $\Gamma_a =\Gamma_b=(1\pm 
\gamma^5)/2$. The function
\begin{flalign}
\Phi_{-4}(x)&= 
\frac{3}{16}
\left(-\frac{K^2_1(x)}{K^2_2(x)} + \frac{K^2_3(x)}{K^2_2(x)}\right)\\
&\sim \frac{3}{2x} +\mathcal{O}(x^{-2})\;\;x\gg 1,\nonumber
\end{flalign}
gives the pure $P$-wave behavior in the nonrelativistic limit. 
The cross section (\ref{sigmatot_k}) behaves as
$\sigma_\text{nr} v_r \sim \sigma_\Lambda a \frac{v^2_r}{4}+O(v^4_r)$.
Using (\ref{Maxwell_vrel}) with reduced mass $\mu=\frac{m}{2}$ we have the average
$\langle v^2_r \rangle_P=\frac{6}{x}$, thus 
$\langle \sigma_\text{nr} v_r \rangle_P=a\sigma_\Lambda  \frac{3}{2x} $ in agreement 
with (\ref{Pphi}).
Further examples are given in \cite{Cannoni:2015wba}.

\subsection{Boltzmann equation}

Let  $\texttt{f}(\boldsymbol{x},\boldsymbol{p},t)$ be a non-equilibrium one particle phase space 
distribution.
In order to shorten the notation we assume now that the factor $g/(2\pi^3)$ is included in 
$\texttt{f}$
and indicate $\texttt{f}_i \equiv \texttt{f}(\boldsymbol{x}_i,\boldsymbol{p}_i,t)$.
The relativistic Boltzmann equation without external forces for binary collisions $1+2\to 3+4$ is 
usually written 
in the invariant form \cite{groot1980relativistic}, \cite{cercignani2002relativistic}
\begin{flalign}
p_1 \cdot \partial\texttt{f}_1=\int\frac{d^3 \boldsymbol{p}_2}{E_2} (\texttt{f}_{3} \texttt{f}_{4} 
-\texttt{f}_1 \texttt{f}_2)
\sigma \mathcal{F},
\label{BEinvariant}
\end{flalign}
where $\partial=(\partial/\partial t, -\nabla)$.
Taking the scalar product on the left-hand side, dividing both sides by $E_1$ and 
remembering that $\mathcal{F}=p_1\cdot p_2 v_\text{rel}$ we obtain
\begin{flalign}
\frac{\partial\texttt{f}_1}{\partial t}+\boldsymbol{v}_1\cdot \nabla\texttt{f}_1 =
\int d^3 \boldsymbol{p}_2 \frac{p_1\cdot p_2}{E_1 E_2} (\texttt{f}_{3} \texttt{f}_{4} 
-\texttt{f}_1 \texttt{f}_2)\sigma v_{\text{rel}}.
\label{BEexplicit}
\end{flalign}
The hyperbolic correlation factor $k_\text{r}$ thus appears explicitly in the collision 
integral.\footnote{
After \cite{groot1980relativistic} it has become popular 
to use the `M\o{}ller velocity'  such that 
the equation takes the form similar to the nonrelativistic  one with $ \bar{v}$ replacing $ v_r$. 
On the other hand in \cite{cercignani2002relativistic} the `M\o{}ller velocity' is introduced
only because simplifies the formulas. 
In our opinion this simplification is deceptive because, as we have seen,
the non physical velocity $\bar{v}$ is source of confusion and errors and hides
the hyperbolic nature of special relativity.} 
\begin{figure}
\includegraphics*[scale=0.7]{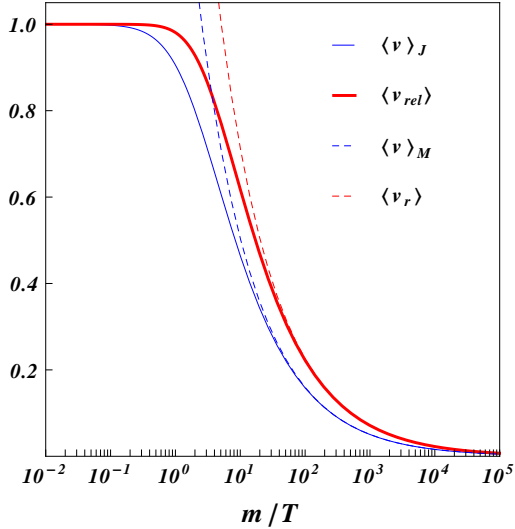}
\caption{Mean values of relative velocity and single particle velocity
given by Eqs.~(\ref{vmean_rel}) and (\ref{vmean_single}).}
\label{Fig:3}
\end{figure}

The average relative velocity (\ref{Vmean_x1_x2}) is useful in 
the relaxation time approximation of the Boltzmann equation 
\cite{AndersonWitting}, \cite{cercignani2002relativistic} 
\begin{flalign}
p\cdot \partial\texttt{f}=-\frac{E}{\tau_\text{coll}}(\texttt{f}-f^\text{eq}),
\end{flalign}
where $\tau_\text{coll}$ is a typical collision time. If the cross section does not vary strongly 
with the relative velocity, we can approximate
$\langle \sigma v_\text{rel}\rangle_\mathcal{P}
\simeq \sigma \langle v_\text{rel} \rangle_\mathcal{P}$ such that 
\begin{flalign}
\tau_\text{coll}\simeq \frac{1}{n^\text{eq} \sigma \langle v_\text{rel} \rangle}_\mathcal{P}.
\end{flalign}
For a gas composed of a single specie of mass $m$ the mean value (\ref{Vmean_x1_x2}) 
is given by the simple expression
\begin{flalign}
\langle v_\text{rel}\rangle_\mathcal{P}
=\frac{4}{x}\frac{K_3(2x)}{K^2_2(x)}
\overset{\text{\tiny $x \gg 1$}}{\longrightarrow}
 \langle v_r\rangle =\sqrt{\frac{16}{\pi x}}.
\label{vmean_rel}
\end{flalign}
This can be confronted for example with the complicated expression 
for $\langle \bar{v}\rangle$ calculated in \cite{cercignani2002relativistic} that
in the ultrarelativistic limit tends to $4/5$, which evidently has no particular 
physical meaning. Another possibility is to use
the mean value of the single particle velocity 
$\langle v\rangle_J=\int d^3 \boldsymbol{p} f_p \frac{|\boldsymbol{p}|}{E}$,
which is easily find to be
\begin{flalign}
\langle v\rangle_J=2\frac{(1+x)}{x^2}\frac{e^{-x}}{K_2(x)}
\overset{\text{\tiny $x \gg 1$}}{\longrightarrow}
\langle v\rangle_M =\sqrt{\frac{8}{\pi x}}.
\label{vmean_single}
\end{flalign}
The averaged velocities and their nonrelativistic expression are shown
in Figure \ref{Fig:3}. Note that $\langle v_\text{rel} \rangle$ and $\langle v\rangle_J$ tend to 1 
in the ultrarelativistic limit $x\ll 1$ as required by 
relativity principle and that the relation $\langle v_r\rangle=\sqrt{2}\langle v\rangle_M$ 
is recovered in the nonrelativistic limit, 
$\langle v_\text{rel}\rangle /\langle v\rangle_J\sim \sqrt{2}+\mathcal{O}(x^{-1})$.

\subsection{Rate equation}

The thermal averaged cross section appears in the integrated Boltzmann equation
that gives the variation in time of the number density of a particle specie
as a consequence of inelastic reactions and expansion of volume.
The system departs
from  chemical equilibrium until it reaches a new stationary state
or moves from a nonequilibrium state towards the equilibrium state. 

An example of the former process is 
the departure from the equilibrium and freeze out of particle species in the 
early Universe.\footnote{See for example \cite{Bernstein:1988bw}, \cite{kolb1994early}, 
\cite{Griest:1990kh}, \cite{Gondolo:1990dk}
for the decoupling of dark matter and other relics.
For the baryogenesis problem see for example \cite{Kolb:1979qa}, \cite{Claudson:1983js}, 
\cite{Luty:1992un}. The use of chemical rate equations
for reaction in the expanding universe was pioneered in \cite{Alpher:1953zz}, \cite{Zeldovich:1966},
\cite{Vysotsky:1977pe}, \cite{Lee77}. } 

An example of the latter is the equilibration of the hadronic system produced 
in high energy nucleus-nucleus collisions after the formation of the quark-gluon plasma.\footnote{
See for example \cite{Biro:1981es},
\cite{Matsui:1985eu}, 
\cite{letessier2002hadrons} and \cite{Schade:2007ex}, \cite{Satarov:2013wga} 
for the connection between the big and the little  bang.}

Let us consider the case of 
the homogeneous and isotropic standard model of cosmology governed by the 
Friedman-Robertson-Walker metric 
with scale factor $a$ and Hubble parameter $H=\frac{1}{a}\frac{da}{dt}$. 
The distribution function depends only on $E$, $|\boldsymbol{p}|$, $t$ and 
the Boltzmann equation takes the form \cite{Bernstein:1988bw}
\begin{flalign}
\frac{\partial\texttt{f}_1}{\partial t} -H \frac{|\boldsymbol{p}_1|}{E_1} \frac{\partial 
\texttt{f}_1}{\partial |\boldsymbol{p}_1|}=
\int d^3 \boldsymbol{p}_2 k_\text{r}
(\texttt{f}_{3} \texttt{f}_{4} -\texttt{f}_1 \texttt{f}_2)\sigma v_{\text{rel}},
\label{BE_cosmo}
\end{flalign}
where we used (\ref{BEexplicit}) to express the collision integral.

Assume that the product particles remain  in equilibrium with zero chemical potential, then
\begin{flalign}
\texttt{f}_{3} \texttt{f}_{4}= f^\text{eq}_3 f^\text{eq}_4=f^\text{eq}_1 f^\text{eq}_2
=n^\text{eq}_1 n^\text{eq}_2 f_{p,1} f_{p,2},
\nonumber
\end{flalign}
where the second equality follows from energy conservation.
Further assume that the  nonequilibrium distribution of the colliding particles 
is proportional to the equilibrium one through the fugacity
$\lambda=\exp(\mu/T)$ where $\mu$ is the chemical potential,
\begin{flalign}
\texttt{f}_{1} \texttt{f}_{2}=\lambda_1 \lambda_2 f^\text{eq}_1 f^\text{eq}_2
=\lambda_1 \lambda_2n^\text{eq}_1 n^\text{eq}_2 f_{p,1} f_{p,2}
=n_1 n_2 f_{p,1} f_{p,2}.
\nonumber
\end{flalign}   
Integrating in $ \boldsymbol{p}_1$ the collision integral in (\ref{BE_cosmo}) we have
\begin{flalign}
(n^\text{eq}_{1} n^\text{eq}_{2} - n_1 n_2)
\int d^3 \boldsymbol{p}_1 d^3 \boldsymbol{p}_2 k_\text{r}
f_{p,1} f_{p,2}\sigma v_{\text{rel}},
\nonumber
\end{flalign}
where the integral is nothing but $\langle \sigma v_{\text{rel}}\rangle_\mathcal{P}$.  
The integral of the left-hand side of (\ref{BE_cosmo}) gives $\frac{dn_1}{dt}+3Hn_1$, thus
\begin{flalign}
\frac{1}{\nu_1} \frac{1}{a^3}\frac{d(n_1 a^3)}{dt}=\frac{1}{j_{12}}(n^\text{eq}_{1} n^\text{eq}_{2} -n_1 n_2)\langle \sigma v_{\text{rel}}\rangle_\mathcal{P}.
\end{flalign}
We have inserted the stoichiometric coefficient \cite{Cannoni:2014zqa} on the left hand side and
the statistical factor on the right hand side.
In this way, when the species 1 and 2 are the same, the equation becomes
\begin{flalign}
\frac{1}{a^3}\frac{d(n a^3)}{dt}=(n^2_\text{eq} -n^2)
\langle \sigma v_{\text{rel}}\rangle_\mathcal{P} ,
\end{flalign}
with $j_{12}=2$  cancelled by stoichiometric coefficient $\nu_1 =2$. For an approach
to rate equations based on nonequilibrium thermodynamics see \cite{Cannoni:2014zqa}.

\section{Final comments }

The Lorentz invariant relative velocity given by formula (\ref{v_rel_Rel_def})
is  practically unknown in quantum field theory and particle physics literature.

We have seen that thanks to (\ref{v_rel_Rel_def}), the formulation of the invariance flux, cross 
section and luminosity becomes simple and transparent. 
In this way the use of non physical velocity like the 
`M\o{}ller velocity', and assumptions about reference frames and collinearity can be avoided.
The cross sections for $ 2 \to 2 $ processes are easily expressed in the rest frame of a particle 
or in the center of momentum frame in terms the invariant relative velocity that also allows to 
obtain the correct nonrelativistic expansion. 

Finally we have reviewed the statistical properties in a relativistic gas
highlighting the role the probability density function of the relative velocity in the determination
of reaction rates, Boltzmann equation and rate equations.
 
The relative velocity determines the   
metric and hyperbolic properties of the velocity space. This is not
a mathematical curiosity but explains the peculiarities 
of the various ingredients that are necessary in formulating the theory of relativistic 
scattering.

\begin{acknowledgments} 

The author acknowledges M. E. G\'{o}mez and J. Rodriguez Quintero.  
A. Pich, G. Rodrigo, J. Portol\'es are acknowledged for hospitality at IFIC in 
Valencia and O. Panella for hospitality at INFN in Perugia
where part of this work was done.
Work supported by
the MINECO grants FPA2011-23778 and FPA2014-53631,
by the grant MULTIDARK CSD2209-00064 of MICINN Consolider-Ingenio 2010 Program,
and by INFN project QU$\_$ASAP. 
\end{acknowledgments}

\bibliography{biblio_review}

\newcommand{\etalchar}[1]{$^{#1}$}
\begin{thebibliography}{IMM{\etalchar{+}}09}

\bibitem[A{\etalchar{+}}11]{Adelberger:2010qa}
E.~G. Adelberger et~al.
\newblock {Solar fusion cross sections II: the pp chain and CNO cycles}.
\newblock {\em Rev. Mod. Phys.}, 83:195, 2011.
\newblock [arXiv:1004.2318].

\bibitem[AFH53]{Alpher:1953zz}
R.~A. Alpher, J.~W. Follin, and R.~C. Herman.
\newblock {Physical conditions in the initial stages of the expanding
  Universe}.
\newblock {\em Phys. Rev.}, 92:1347, 1953.

\bibitem[AW74]{AndersonWitting}
J.~L. Anderson and H.~R. Witting.
\newblock {A relativistic relaxation time model for the Boltzmann equation}.
\newblock {\em Physica}, 74:466, 1974.

\bibitem[Bar11]{Barrett:2011}
J.~F Barrett.
\newblock {The hyperbolic theory of special relativity}.
\newblock arXiv:1102.0462, 2011.

\bibitem[BBLZ83]{Biro:1981es}
T.~Biro, H.~W. Barz, B.~Lukacs, and J.~Zimanyi.
\newblock {Entropy and hadrochemical composition in heavy ion collision}.
\newblock {\em Phys. Rev. C}, 27:2695, 1983.

\bibitem[BD64]{bjorken1964relativistic}
J.~D. Bjorken and S.~D. Drell.
\newblock {\em {Relativistic quantum mechanics}}.
\newblock McGraw-Hill, New York, 1964.

\bibitem[Ber88]{Bernstein:1988bw}
J.~Bernstein.
\newblock {\em {Kinetic theory in the expanding Universe}}.
\newblock Cambridge University, New York, 1988.

\bibitem[BLP82]{berestetskii2012quantum}
V.~B. Berestetskii, E.~M. Lifshitz, and L.~P. Pitaevskii.
\newblock {\em {Quantum Electrodynamics (2nd ed.)}}.
\newblock Butterworth-Heinemann, Oxford, 1982.

\bibitem[Bro94]{Brown:1992db}
L.~S. Brown.
\newblock {\em {Quantum field theory}}.
\newblock Cambridge University, Cambridge, 1994.

\bibitem[Can14]{Cannoni:2013bza}
M.~Cannoni.
\newblock {Relativistic $\langle\sigma v_\text{rel}\rangle$ in the calculation
  of relics abundances: a closer look}.
\newblock {\em Phys. Rev. D}, 89:103533, 2014.
\newblock [arXiv:1311.4494, 1311.4508].

\bibitem[Can15]{Cannoni:2014zqa}
M.~Cannoni.
\newblock {Exact theory of freeze out}.
\newblock {\em Eur. Phys. J. C}, 75:106, 2015.
\newblock [arXiv:1407.4108].

\bibitem[Can16]{Cannoni:2015wba}
M.~Cannoni.
\newblock {Relativistic and nonrelativistic annihilation of dark matter: a
  sanity check using an effective field theory approach}.
\newblock {\em Eur. Phys. J. C}, 76:137, 2016.
\newblock [arXiv:1506.07475].

\bibitem[CHH84]{Claudson:1983js}
M.~Claudson, L.~J. Hall, and I.~Hinchliffe.
\newblock {Cosmological baryon generation at low temperatures}.
\newblock {\em Nucl. Phys. B}, 241:309, 1984.

\bibitem[CK02]{cercignani2002relativistic}
C.~Cercignani and G.~M. Kremer.
\newblock {\em {The relativistic Boltzmann equation: Theory and applications}}.
\newblock Birkh{\"a}user, Basel, 2002.

\bibitem[Ein05]{Einstein:1905ve}
A.~Einstein.
\newblock {On the electrodynamics of moving bodies}.
\newblock {\em Annalen Phys.}, 17:891, 1905.

\bibitem[Foc64]{fock1969theory}
V.~A. Fock.
\newblock {\em {The theory of space, time and gravitation (2nd ed.)}}.
\newblock Pergamon, Oxford, 1964.

\bibitem[Fur03]{Furman}
F.~A. Furman.
\newblock {The M\o{}ller luminosity factor}.
\newblock LBNL-53553, 2003.

\bibitem[GG91]{Gondolo:1990dk}
P.~Gondolo and G.~Gelmini.
\newblock {Cosmic abundances of stable particles: Improved analysis}.
\newblock {\em Nucl. Phys. B}, 360:145, 1991.

\bibitem[GLvW80]{groot1980relativistic}
S.~R. Groot, W.~A. Leeuwen, and C.~G. van Weert.
\newblock {\em {Relativistic kinetic theory: Principles and applications}}.
\newblock North-Holland, Amsterdam, 1980.

\bibitem[Gou13]{gourgoulhon2013special}
E.~Gourgoulhon.
\newblock {\em {Special relativity in general frames: from particles to
  astrophysics}}.
\newblock Springer, Berlin, 2013.

\bibitem[GR08]{greiner2008quantum}
W.~Greiner and J.~Reinhardt.
\newblock {\em {Quantum electrodynamics}}.
\newblock Springer, Berlin, 2008.

\bibitem[GS91]{Griest:1990kh}
K.~Griest and D.~Seckel.
\newblock {Three exceptions in the calculation of relic abundances}.
\newblock {\em Phys. Rev. D}, 43:3191, 1991.

\bibitem[GW64]{goldberger1964collision}
M.~L. Goldberger and K.~M. Watson.
\newblock {\em {Collision theory}}.
\newblock Wiley, New York, 1964.

\bibitem[Hag73]{hagedorn1973relativistic}
R.~Hagedorn.
\newblock {\em Relativistic kinematics}.
\newblock Benjamin, New York, 1973.

\bibitem[HM84]{Halzen:1984mc}
F.~Halzen and A.~D. Martin.
\newblock {\em {Quarks and Leptons: an introductory course in modern particle
  physics}}.
\newblock Wiley, New York, 1984.

\bibitem[HM03]{Herr:2003em}
W.~Herr and B.~Muratori.
\newblock {Concept of luminosity}.
\newblock In {\em Proceedings of CERN Accelerator School, Zeuthen, Germany,
  September 15-26, 2003}, page 361, 2003.

\bibitem[IMM{\etalchar{+}}09]{Iocco:2008va}
F.~Iocco, G.~Mangano, G.~Miele, O.~Pisanti, and P.~D. Serpico.
\newblock {Primordial nucleosynthesis: from precision cosmology to fundamental
  physics}.
\newblock {\em Phys. Rept.}, 472:1, 2009.
\newblock [arXiv:0809.0631].

\bibitem[JR55]{jauch1955theory}
J.~M. Jauch and F.~Rohrlich.
\newblock {\em {The theory of photons and electrons}}.
\newblock Addison-Wesley, Reading, 1955.

\bibitem[Kak93]{Kaku:1993ym}
M.~Kaku.
\newblock {\em {Quantum field theory: A modern introduction}}.
\newblock Oxford, New York, 1993.

\bibitem[KT90]{kolb1994early}
E.~Kolb and M.~Turner.
\newblock {\em {The early Universe}}.
\newblock Addison-Wesley, Reading, 1990.

\bibitem[KW80]{Kolb:1979qa}
E.~W. Kolb and S.~Wolfram.
\newblock {Baryon number generation in the early Universe}.
\newblock {\em Nucl. Phys. B}, 172:224, 1980.
\newblock [Erratum: Nucl. Phys. B 195, 5421982].

\bibitem[LL75]{Landau:1982dva}
L.~D. Landau and E.~M. Lifschitz.
\newblock {\em {The classical theory of fields (4th ed.)}}.
\newblock Butterworth-Heinemann, Oxford, 1975.

\bibitem[LL80]{landau2013statistical}
L.~D. Landau and E.~M. Lifshitz.
\newblock {\em {Statistical Physics (3rd ed.)}}.
\newblock Butterworth-Heinemann, Oxford, 1980.

\bibitem[LR02]{letessier2002hadrons}
J.~Letessier and J.~Rafelski.
\newblock {\em {Hadrons and quark--gluon plasma}}.
\newblock Cambridge University, Cambridge, 2002.

\bibitem[Lut92]{Luty:1992un}
M.~A. Luty.
\newblock {Baryogenesis via leptogenesis}.
\newblock {\em Phys. Rev. D}, 45:455, 1992.

\bibitem[LW77]{Lee77}
B.~W. Lee and S.~Weinberg.
\newblock Cosmological lower bound on heavy neutrino masses.
\newblock {\em Phys. Rev. Lett.}, 39:165, 1977.

\bibitem[M\o45]{Moller1945}
C.~M\o{}ller.
\newblock {General properties of the characteristic matrix in the theory of
  elementary particles}.
\newblock {\em D. Kgl Danske Vidensk. Selsk. Mat.-Fys. Medd.}, 23:1, 1945.

\bibitem[MS63]{Middelkoop:1963zz}
W.~C. Middelkoop and A.~Schoch.
\newblock {Interaction rate in colliding beams systems}.
\newblock CERN-AR-SG-63-40, 1963.

\bibitem[MSM86]{Matsui:1985eu}
T.~Matsui, B.~Svetitsky, and L.~D. McLerran.
\newblock {Strangeness production in ultrarelativistic heavy ion collisions. 1.
  Chemical kinetics in the quark--gluon plasma}.
\newblock {\em Phys. Rev. D}, 34:783, 1986.
\newblock [Erratum: Phys. Rev.D {\bf 37}, 844 (1988)].

\bibitem[Nap93]{Napoly:1992kn}
O.~Napoly.
\newblock {The luminosity for beam distributions with error and wake field
  effects in linear colliders}.
\newblock {\em Part. Accel.}, 40:181, 1993.

\bibitem[O{\etalchar{+}}14]{Agashe:2014kda}
K.~A. Olive et~al.
\newblock {Review of Particle Physics}.
\newblock {\em Chin. Phys.}, C 38:090001, 2014.

\bibitem[PS95]{Peskin:1995ev}
M.~E. Peskin and D.~V. Schroeder.
\newblock {\em {An introduction to quantum field theory}}.
\newblock Addison-Wesley, Reading, 1995.

\bibitem[RS04]{Rhodes:2004}
J.~A. Rhodes and M.~D. Semon.
\newblock {Relativistic velocity space, Wigner rotation, and Thomas
  precession}.
\newblock {\em Am. J. Phys.}, 72:943, 2004.

\bibitem[SK09]{Schade:2007ex}
H.~Schade and B.~Kampfer.
\newblock {Antiproton evolution in little bangs and big bang}.
\newblock {\em Phys. Rev. C}, 79:044909, 2009.
\newblock [arXiv:0705.2003].

\bibitem[SMG13]{Satarov:2013wga}
L.~M. Satarov, I.~N. Mishustin, and W.~Greiner.
\newblock {Evolution of antibaryon abundances in the early Universe and in
  heavy-ion collisions}.
\newblock {\em Phys. Rev. C}, 88:024908, 2013.
\newblock [arXiv:1305.4046].

\bibitem[Ter70]{Terrall}
J.~R. Terrall.
\newblock {Elementary treatment of relativistic cross sections}.
\newblock {\em Am. J. Phys.}, 38:1460, 1970.

\bibitem[Tsa10]{tsamparlis2010special}
M.~Tsamparlis.
\newblock {\em {Special relativity: An Introduction with 200 problems and
  solutions}}.
\newblock Springer, Berlin, 2010.

\bibitem[Tul11]{Tully:2011zz}
C.~G. Tully.
\newblock {\em {Elementary particle physics in a nutshell}}.
\newblock Princeton University, Princeton, 2011.

\bibitem[VDZ77]{Vysotsky:1977pe}
M.~I. Vysotsky, A.~D. Dolgov, and {\relax Ya}.~B. Zeldovich.
\newblock {Cosmological restriction on neutral lepton masses}.
\newblock {\em JETP Lett.}, 26:188, 1977.
\newblock [Pisma Zh. Eksp. Teor. Fiz. {\bf 26}, 200 (1977)].

\bibitem[Wea76]{Weaver}
T.~A. Weaver.
\newblock {Reaction rates in a relativistic plasma}.
\newblock {\em Phys. Rev. A}, 13:1563, 1976.

\bibitem[Wei95]{Weinberg:1995mt}
S.~Weinberg.
\newblock {\em {The quantum theory of fields. Vol. 1: Foundations}}.
\newblock Cambridge University, Cambridge, 1995.

\bibitem[Zee10]{Zee:2003mt}
A.~Zee.
\newblock {\em {Quantum field theory in a nutshell, (2nd Ed.)}}.
\newblock Princeton University, Princeton, 2010.

\bibitem[ZOP66]{Zeldovich:1966}
Ya.~B. Zel’dovich, L.~B. Okun, and S.~B. Pikel’ner.
\newblock Quarks: astrophysical and physicochemical aspects.
\newblock {\em Sov. Phys. Usp.}, 8:702, 1966.
\newblock [Usp. Fiz. Nauk 87, 113 (1965)].

\end{thebibliography}

\end{document}